\documentclass[final,5p,twocolumn]{elsarticle}

\usepackage{lineno}
\usepackage{amsmath}
\usepackage[colorlinks]{hyperref}
\usepackage{natbib}
\usepackage{geometry}
\usepackage{fleqn}
\usepackage{graphicx}
\usepackage{txfonts}
\usepackage{xspace}
\usepackage{xcolor}

\newcommand{\be}{\begin{equation}}
\newcommand{\ee}{\end{equation}}
\newcommand{\Eq}[1]{Eq.~\eqref{#1}}
\newcommand{\Eqs}[2]{Eqs.~\eqref{#1} and \eqref{#2}}

\newcommand{\Fig}[1]{Fig.~\ref{#1}}
\newcommand{\Ref}[1]{Ref.~\cite{#1}}

\newcommand{\dbd}[2]{\frac{\mathrm{d}#1}{\mathrm{d}#2}}

\newcommand{\runDM}{\textsc{runDM}\xspace}
\newcommand{\LambdaUV}{\Lambda_{\rm UV}}

\biboptions{sort&compress}

\begin{document}

\begin{frontmatter}

\title{Probing Leptophilic Dark Sectors with Hadronic Processes}

\author[UCSC,SCIPP]{Francesco D'Eramo}
\ead{fderamo@ucsc.edu}

\author[Jussieu]{Bradley J. Kavanagh}
\ead{bradley.kavanagh@lpthe.jussieu.fr}

\author[CERN,IAP]{Paolo Panci}
\ead{paolo.panci@cern.ch}

\address[UCSC]{Department of Physics, University of California Santa Cruz, 1156 High St., Santa Cruz, CA 95064, USA}
\address[SCIPP]{Santa Cruz Institute for Particle Physics, 1156 High St., Santa Cruz, CA 95064, USA}
\address[Jussieu]{Laboratoire de Physique Th\'{e}orique et Hautes Energies, CNRS, UMR 7589, 4 Place Jussieu, F-75252, Paris, France}
\address[CERN]{CERN Theoretical Physics Department, CERN, Case C01600, CH-1211 Gen\`eve, Switzerland}
\address[IAP]{Institut d'Astrophysique de Paris, UMR 7095 CNRS, Universit\'e Pierre et Marie Curie,
98 bis Boulevard Arago, Paris 75014, France}

\begin{abstract}
We study vector portal dark matter models where the mediator couples only to leptons. In spite of the lack of tree-level couplings to colored states, radiative effects generate interactions with quark fields that could give rise to a signal in current and future experiments. We identify such experimental signatures: scattering of nuclei in dark matter direct detection; resonant production of lepton-antilepton pairs at the Large Hadron Collider; and hadronic final states in dark matter indirect searches. Furthermore, radiative effects also generate an irreducible mass mixing between the vector mediator and the $Z$ boson, severely bounded by ElectroWeak Precision Tests. We use current experimental results to put bounds on this class of models, accounting for both radiatively induced and tree-level processes. Remarkably, the former often overwhelm the latter.
\end{abstract}


\end{frontmatter}

\section{Introduction}

Weakly Interacting Massive Particles (WIMPs) are motivated dark matter (DM) candidates testable with multiple and complementary methods~\cite{Jungman:1995df,Bertone:2004pz,Feng:2010gw}. An attractive class of WIMP models consists of those in which leptons are the only Standard Model (SM) degrees of freedom coupled to the DM~\cite{Fox:2008kb,Kopp:2009et,Bell:2014tta,Freitas:2014jla,delAguila:2014soa,Chen:2015tia}. Much interest in {\it leptophilic} models was ignited by the excess in the positron fraction at high energy observed by the  PAMELA experiment~\cite{Adriani:2008zr,Adriani:2013uda}, later confirmed by Fermi~\cite{FermiLAT:2011ab} and AMS-02~\cite{Aguilar:2013qda,Accardo:2014lma}. No associated excess in the anti-proton flux has been observed. DM particles annihilating to leptons can provide a good fit to the positron excess, although complementary bounds from gamma-rays (see e.g.~\Ref{Cirelli:2009dv,Meade:2009iu,FermiLAT:2012bf}) and Cosmic Microwave Background (see e.g.~\Ref{Madhavacheril:2013cna,Slatyer:2015jla}) challenge this hypothesis and astrophysical explanations for this excess (e.g.~pulsars~\cite{Hooper:2008kg,Profumo:2008ms,Feng:2015uta}) remain plausible.

The expected experimental signals for leptophilic models are quite distinct. The absence of couplings to quarks and gluons leads (na\"ively) to vanishing rates for direct detection and for DM production at hadron colliders. Furthermore, DM annihilation would lead to indirect detection spectra from only lepton final states. For these reasons nearly all phenomenological studies have focused on bounds coming from tree-level processes. These include mono-photon and lepton production at lepton colliders, four-lepton events at hadron colliders and diffuse gamma rays from DM annihilations. There are two noteworthy exceptions. On one hand, loop-induced contributions to DM scattering of target nuclei give rates observable in direct detection, as first pointed out by \Ref{Kopp:2009et}. On the other hand, loop diagrams with a virtual leptophilic mediator give substantial corrections to lepton anomalous magnetic moments~\cite{Agrawal:2014ufa}.

In this work, we identify new loop-induced signals coming from leptophilic dark sectors. We focus on the broad class of models where interactions between DM and leptons are mediated by a heavy vector particle. Radiative effects are accounted for by solving the renormalization group (RG) equations describing the evolution of the couplings with the energy scale~\cite{Crivellin:2014qxa,D'Eramo:2014aba}, a procedure recently automated with the public code \runDM~\cite{D'Eramo:2016atc,runDM}. We use these tools to explore the phenomenology of leptophilic models by rigorously accounting for the different energy scales.  Crucially, different search strategies probe dark sector couplings at different energy scales, and RG flow induces new couplings through operator mixing. Despite the radiatively-induced couplings being suppressed by at least one loop factor, they can still give significant bounds, particularly in the case of loop-induced couplings to quarks (for related studied of RG effects see Refs.~\cite{Hill:2011be,Frandsen:2012db,Haisch:2013uaa,Hill:2013hoa,Kopp:2014tsa,Crivellin:2014gpa,Hill:2014yxa,D'Eramo:2016mgv}).

Our work extends the previous literature on vector portal leptophilic models by identifying new processes arising from such loop-induced couplings and imposing the associated experimental constraints. As a result, we get novel bounds on the parameter space from the following signals:
\begin{itemize}
\item \textbf{LEP-II compositeness bound for $\mu$ and/or $\tau$.} For models where the mediator couples only to $\mu$ and/or $\tau$, we evaluate the RG-induced contribution to the coupling between the mediator and the electron. We constrain it by using LEP-II data for production of leptons in the final state.
\item \textbf{Mass mixing with the $Z$ boson.} We identify the RG-induced contribution to the mass mixing between the leptophilic vector mediator and the $Z$ boson. We then place bounds using ElectroWeak Precision Tests (EWPT) data.
\item \textbf{Dilepton resonance at the LHC.} We find the RG-induced contributions to the coupling between the mediator and light quarks. We constrain this coupling using data from LHC searches for dilepton resonances. 
\item \textbf{Direct detection.} Still using the RG-induced couplings between the mediator and light quarks, but this time at a much lower energy, we compute the DM elastic scattering rate of target nuclei and we impose direct detection bounds. If the mediator couples to the DM vector current, we improve the result of \Ref{Kopp:2009et} by applying the RG analysis developed in \Ref{D'Eramo:2014aba}. On the contrary, if the mediator couples to the DM axial-vector current, we point out a contribution to the scattering rate proportional to the Yukawa coupling of the lepton under consideration. This effect was first pointed out in \Ref{D'Eramo:2014aba}, and has not appeared in any previous analysis of leptophilic DM.
\end{itemize}
As a remarkable result of our analysis, we show that the most stringent constraints come from the RG-induced processes described above in a large region of the parameter space. 

We describe in Sec.~\ref{sec:simplifiedmodel} the simplified model framework we adopt in this study. After setting up the notation, we give a simplified analytical solution for the RG evolution of the couplings in Sec.~\ref{sec:RGE}. While the analysis in this work is performed by employing the code \runDM, these analytical solutions correctly capture the order of magnitude size of the the full numerical results. The experimental constraints are enumerated in Sec.~\ref{sec:constraints}, where we highlight which bounds are present at the tree-level and which arise from RG-induced couplings. We present these bounds in Fig.~\ref{fig:Limits} and we compare them for couplings to different lepton flavors in Sec.~\ref{sec:results}. In Sec.~\ref{sec:contamination}, we discuss in more detail the possible hadronic contamination of indirect detection spectra, an effect which has not previously been pointed out within this class of models. Finally, we present our conclusions in Sec.~\ref{sec:conclusions}.

\section{A Leptophilic Vector Mediator}
\label{sec:simplifiedmodel}

We augment the field content of the SM with two singlets: a fermion DM candidate $\chi$ assumed to be stabilized by a $Z_2$ symmetry, and a vector mediator $V$. We do not specify the microscopic origin of the mediator, but rather we employ a simplified model framework with interaction
\be
\mathcal{L}_{\rm int} = \left(J_\mu^{\rm (DM)} + J_\mu^{\rm (ch. \, leptons)} + J_\mu^{\rm (neutrinos)} \right)  V^\mu  \ .
\label{eq:LagSimplifiedModel}
\ee
First, the mediator has tree-level couplings with the DM field through the fermion current
\be
J_\mu^{\rm (DM)} = \overline{\chi} \, \gamma_\mu \left( g_{V \chi} + g_{A \chi} \gamma^5 \right) \chi \ .
\label{eq:JmuDM}
\ee
We perform our phenomenological study for a Dirac DM field. The analysis for a Majorana DM particle would be very similar, with the absence of the DM vector current in \Eq{eq:JmuDM} and appropriate factors of $2$ in different observables. We find it convenient to parameterize the charged lepton current in terms of the vector and axial-vector pieces
\be
J_\mu^{\rm (ch. \, leptons)} =  \sum_{l = e, \mu, \tau} \overline{l} \, \gamma_\mu \left(g_{V l} + g_{A l} \gamma^5 \right) l \ .
\label{eq:Jmuleptons} 
\ee
Although not electroweak gauge invariant, the connection with gauge invariant currents is straightforward. A manifestly gauge invariant lepton current would have two separate terms, associated to the left-handed weak isospin doublet $l_L$ and the right-handed weak isospin singlet $e_R$. These gauge invariant currents can be unambiguosly translated into the ones in \Eq{eq:Jmuleptons} by using the relation $g_{V,A \, l} = (\pm c_{l_L} + c_{e_R}) / 2$. As a consequence of electroweak gauge invariance, the vector mediator has to couple also to SM neutrinos
\be
J_\mu^{\rm (neutrinos)} =  \sum_{l = e, \mu, \tau} \left( \frac{g_{V l} - g_{A l}}{2} \right) \; \overline{\nu_l} \, \gamma_\mu \left(1 - \gamma^5 \right) \nu_l \ .
\label{eq:Jmuneutrinos} 
\ee
The only way to avoid neutrino couplings is for a mediator coupled to both vector and axial-vector currents of charged leptons with identical coefficient. This happens if the mediator couples only to the right-handed weak isospin singlet $e_R$, which is why the couplings to neutrinos vanish in this limit. 

The simplified model with interactions as in \Eq{eq:LagSimplifiedModel} can only be valid up to some cut-off scale $\Lambda_{\rm UV}$. Above such a scale, we expect new degrees of freedom UV-completing the simplified model and protecting the mediator couplings to other SM fields. The dimensionless couplings in the DM and lepton currents in \Eqs{eq:JmuDM}{eq:Jmuleptons}, and the gauge invariant couplings to neutrino currents in \Eq{eq:Jmuneutrinos}, are always defined at the renormalization scale $\Lambda_{\rm UV}$. RG effects driven by SM couplings, mainly through operator mixing, induce new interactions at energy scales lower than $\Lambda_{\rm UV}$ that can have a phenomenological relevance, as we will extensively discuss in this work. For concreteness, we fix $\Lambda_{\rm UV} = 10 \,\,\mathrm{TeV}$, though typically the induced couplings depend only logarithmically on the cut-off scale, so small changes to this value would not affect our results substantially.

\section{RG Analytical Solutions}
\label{sec:RGE}

The couplings to SM particles other than leptons induced by RG evolutions at energy scales below $\Lambda_{\rm UV}$ are numerically evaluated by using the code \runDM. In this Section, we provide the reader with simple analytical solutions to the RG equations derived in Ref.~\cite{D'Eramo:2014aba}, which are helpful to understand the size of the effects. The analytical results are obtained by a fixed order calculation, whereas \runDM also accounts for the evolution of SM couplings (see App.~A of \Ref{D'Eramo:2016atc}). For leptophilic models such a difference is pretty moderate, hence the solutions given here capture the RG effects with very good accuracy. 

\begin{figure}[t!]
\centering
{\includegraphics[width=0.47\textwidth]{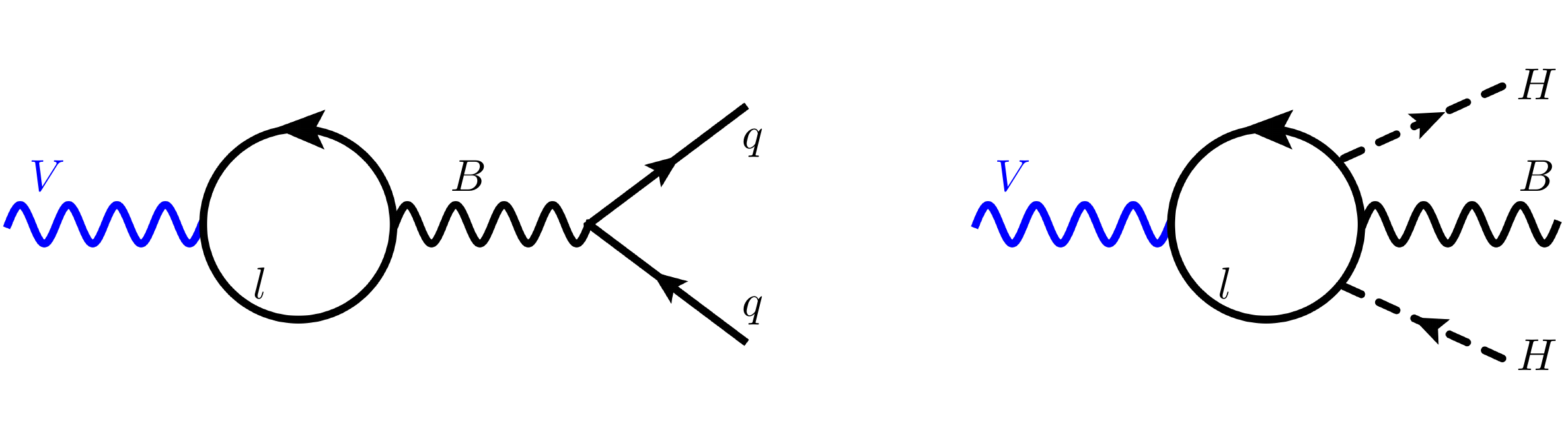}}
\caption{Feynman diagrams for the operator mixing investigated in this work. Left: kinetic mixing between $V$ and the hypercharge gauge boson $B$ (and with the photon below the EWSB scale) induces mediator couplings to quarks. Right: loop-induced coupling between the mediator and the Higgs current that in turn generates mass mixing between $V$ and the $Z$ boson (plus other diagrams with only $H$ external legs and with the SU(2)$_L$ gauge bosons).}
\label{fig:FeynDiag}
\end{figure}

At renormalization scales $\mu$ between the cutoff and the electroweak scale, $\Lambda_{\rm UV} \gtrsim \mu \gtrsim m_Z$, all SM fields are in the spectrum. Radiative corrections driven by hypercharge interactions (Feynman diagram on the left of \Fig{fig:FeynDiag}) induce flavor universal couplings between the mediator and SM fermions
\be
c_i(\mu) = c_i(\LambdaUV) + \frac{2}{3} \frac{\alpha_Y}{\pi} \, y_i \, \left( \sum_{l = e, \mu, \tau} g_{V l}  \right)
\log\left(\Lambda_{\rm UV} / \mu\right) \ .
\label{eq:RGEappendixHypercharge}
\ee
Here, $\alpha_Y = g_Y^2 / (4 \pi)$ the hypercharge fine structure constant. The index $i$ runs over all possible $5$ SM fermions with well defined electroweak quantum numbers and hypercharge $y_i$. Crucially, the result in \Eq{eq:RGEappendixHypercharge} accounts for both charged leptons and neutrinos in the loop, as dictated by the gauge invariant analysis in \Ref{D'Eramo:2014aba}. Using the same notation in Eq.~\eqref{eq:Jmuleptons}, we employ the general result in \Eq{eq:RGEappendixHypercharge} to identify the induced vector and axial-vector currents of up- and down-type quarks
\begin{align}
& \, \left\{g_{V u}(\mu), g_{V d}(\mu), g_{A u}(\mu), g_{A d}(\mu) \right\} =  \\ & \, \nonumber
\left\{ \frac{5}{18}, - \frac{1}{18}, \frac{1}{6} , - \frac{1}{6} \right\} \, \frac{\alpha_Y}{\pi} \, \left( \sum_{l = e, \mu, \tau} g_{V l}  \right)
\log\left(\Lambda_{\rm UV} / \mu\right)  \ ,
\label{eq:onelooptoquarks}
\end{align}
The same Feynman diagrams generate loop-induced currents to leptons as well, relevant only in the absence of the associated tree-level term. Hypercharge interactions generate flavor universal lepton currents
\begin{align}
& \, \left\{\Delta g_{V l}(\mu), \Delta g_{A l}(\mu) \right\}_Y = \\ & \, \nonumber
\left\{- \frac{1}{2}, -  \frac{1}{6} \right\} \, \frac{\alpha_Y}{\pi} \, \left( \sum_{l = e, \mu, \tau} g_{V l}  \right)
\log\left(\Lambda_{\rm UV} / \mu\right)  \ ,
\end{align}
where we use the symbol $\Delta$ to emphasize that these are radiative corrections with respect to the tree-level terms in Eq.~\eqref{eq:Jmuleptons}. The last operator we discuss above the weak scale is the mediator coupling with the Higgs current 
\be
\Delta \mathcal{L}_H \equiv g_H \, H^\dagger i \overleftrightarrow D_\mu H \, V^\mu \ ,
\ee
generated by the Feynman diagram on the right of \Fig{fig:FeynDiag}. Upon neglecting the electron and muon Yukawa couplings, the associated coefficient reads
\be
g_H(\mu) = \left[  \frac{1}{3}  \frac{\alpha_Y}{\pi}  \left(\sum_{l = e, \mu, \tau} g_{V l} \right)
- \frac{\alpha_\tau}{\pi} g_{A \tau} \right] \log\left(\Lambda_{\rm UV} / \mu\right) \ ,
\label{eq:gHapprox}
\ee
where we define $\alpha_\tau = \lambda_\tau^2 / (4 \pi)$ as the effective Yukawa coupling for the $\tau$.

Heavy SM degrees of freedom are integrated-out below the weak scale. Integrating out the Z boson gives rise to threshold corrections to the vector and axial-vector currents of SM fermions coupled to the mediator, as a consequence of the induced $g_H$ in \Eq{eq:gHapprox} (for details of integrating out heavy SM states see the analysis in \Ref{D'Eramo:2014aba}). DD rates are set by the Wilson coefficients of the contact interactions
\be
\mathcal{L}_{\rm DD} = - \frac{J_\mu^{\rm (DM)}}{m_V^2}  
\left[ \sum_{q = u, d} \mathcal{C}_{V}^{(q)} \; \overline{q} \gamma^\mu q + 
\sum_{q = u, d, s} \mathcal{C}_{A}^{(q)} \; \overline{q} \gamma^\mu \gamma^5 q \right] \ .
\label{eq:LagDD}
\ee
Here, we distinguish between coupling to vector and axial-vector currents of light quarks, and we only keep the ones relevant for direct detection rates. The approximate expressions for the Wilson coefficients read
\begin{align}
\label{eq:CVq} \mathcal{C}_{V}^{(q)} =  & \, \frac{2 \alpha_{\rm e.m.}}{3 \pi} \, Q_q \,  \left( \sum_{l = e, \mu, \tau} g_{V l} \right) \,
\log\left(\Lambda_{\rm UV} / \mu_N\right) + \\ & \nonumber
 -  \frac{\alpha_\tau}{\pi} \left(T^{(3)}_q - 2 s_w^2 Q_q \right) g_{A \tau} \log\left(\Lambda_{\rm UV} / \mu_N\right)  \ .
\end{align}
Here, $\alpha_{\rm e.m.} = e^2 / (4 \pi)$ is the electromagnetic fine structure constant, whereas the nuclear scale relevant for DD is $\mu_N \simeq 1 \, {\rm GeV}$. The third component of the weak isospin and the electric charge for the up and down quarks are $(T^{(3)}_u, T^{(3)}_d) = ( + 1/2, -1/2)$ and $(Q_u, Q_d) = ( + 2/3, -1/3)$, respectively. Likewise, RG-induced interactions with quark axial currents read
\be
\mathcal{C}_{A}^{(q)} = T^{(3)}_q \frac{\alpha_\tau}{\pi} g_{A \tau} \log\left(\Lambda_{\rm UV} / \mu_N\right)  \ .
\label{eq:CAq}
\ee

\section{Phenomenological Constraints}
\label{sec:constraints}

We list the experimental bounds on leptophilic models by dividing them into three different categories. First, we consider tree-level processes. An intermediate case is for experimental signals that could be present at the tree-level, depending on which specific lepton flavor is coupled to the mediator. Lastly, we list genuinely loop-induced constraints that cannot be obtained by considering only the operators in \Eq{eq:LagSimplifiedModel}.  All these constraints are implemented in the phenomenological analysis performed in Sec.~\ref{sec:results}, where we consider the mediator coupled to a single lepton flavor at a time. As explicitly shown in Fig.~\ref{fig:Limits}, the loop-induced bounds frequently dominate over the tree-level ones.

\subsection{Bounds from tree-level processes}

\vspace{0.2cm}

\paragraph{Perturbative Unitarity} The simplified model defined in Sec.~\ref{sec:simplifiedmodel} may violate perturbative unitarity at high energies. For pure vector couplings ($g_{Al} = g_{A\chi} = 0$) there is no unitarity issue, since in this case the mediator mass can be generated via a Stueckelberg mechanism~\cite{Stueckelberg:1900zz} without any further low-energy degrees of freedom. Bounds from unitarity arise then only when axial couplings are present, putting a lower bound on the mediator mass~\cite{Kahlhoefer:2015bea}. The leptons are much lighter than the mediator masses considered in this work, so they are massless for our purposes and we do not get any bound in the presence of the axial coupling $g_{Al}$. On the contrary, an axial-vector coupling to DM particles gives a lower bound for the mediator mass
\be
m_V \gtrsim \sqrt{2 / \pi} \, g_{A\chi} \, m_\chi \ .
\label{eq:UnitarityBound}
\ee

\paragraph{Relic Density} We numerically solve the Boltzmann equation for the DM number density, with the thermally averaged cross section computed as prescribed by \Ref{Gondolo:1990dk}. In particular, we fully account for relativistic corrections and for resonance effects, the latter evaluated by computing the mediator width self-consistently with the simplified model field content. The non-relativistic limit for the annihilation cross sections are collected in~\ref{app:XS}. We include in our analysis Sommerfeld corrections~\cite{Hisano:2004ds} to the annihilation cross section. As it turns out, they are relevant only for annihilation to mediators. For a DM particle coupled with the vector current, we use a standard analytic expression of the Sommerfeld factor obtained by replacing the Yukawa with the  Hulth\'en potential (see e.g.~~\cite{Cassel:2009wt,Slatyer:2009vg,Cirelli:2016rnw}). In the phenomenological analysis of Sec.~\ref{sec:results}, we do not need to include such effects for DM coupled with the axial-vector current, since we always fix $g_{A \chi} = 1$; the region where Sommerfeld effects would be relevant ($m_\chi \gtrsim m_V$) is then excluded by perturbative unitarity (see \Eq{eq:UnitarityBound}). 

Constraints from relic density are not treated on an equal footing with the other experimental bounds listed in this Section. The computed dark matter abundance relies upon the extrapolation of the universe snapshot at the time of Big Bang Nucleosynthesis ($T_{\rm BBN} \simeq 1 \, {\rm MeV}$), where the universe was a thermal bath of relativistic particles, to higher temperatures. We have no hint about the energy content of the universe for higher temperatures, relevant for dark matter freeze-out. Thus we remain agnostic about the thermal history at temperatures above $T_{\rm BBN}$, and we always assume that the dark matter is produced with the observed relic abundance when we impose limits such as direct and indirect detection. Our relic density lines serve only as benchmark values. Upon considering modified cosmological histories, a correct relic density can be achieved with annihilation cross sections much smaller~\cite{McDonald:1989jd,Kamionkowski:1990ni,Chung:1998rq,Giudice:2000ex} or much larger~\cite{Profumo:2003hq,Salati:2002md,DEramo:2017gpl} than the thermal relic value. 

\paragraph{Indirect Detection} We impose bounds from observations of diffuse gamma rays from dwarf spheroidal galaxies (dSphs). In the $m_\chi < m_V$ regime, DM can only annihilate to leptons and we impose the limits obtained by the Fermi collaboration in Ref~\cite{Ackermann:2015zua}. The Fermi Collaboration only report limits on the cross section for annihilation into single channels ($e^\pm$, $\mu^\pm$, $\tau^\pm$, etc.). For flavor universal interactions, annihilation proceeds into a mixture of channels. In that case, the limits can be calculated using the publicly available Fermi likelihoods \cite{FermiLikelihood}, combined with the annihilation spectra from \textsc{PPPC4DMID} \cite{Cirelli:2010xx,Ciafaloni:2010ti}.  Once annihilations to mediators open up, we use the constraints of Ref.~\cite{Elor:2015bho} for cascade annihilation into leptons. Given the small DM velocity, the inclusion of Sommerfeld corrections~\cite{Hisano:2004ds} is of importance, and we account for them as in the relic density calculation described above.

In Sec.~\ref{sec:contamination}, we explore the effects which loop-induced couplings (in particular to quarks) may have on annihilation spectra and therefore on indirect detection limits.

\subsection{Bounds that can be present at tree-level}

\vspace{0.2cm}

\paragraph{LEP-II Compositeness Bound} Measurements of cross sections and forward-backward asymmetries for the production of leptons at LEP-II have been used to set limits on possible 4-fermion contact interactions and, in turn, the mass of a new vector mediator which could mediate such interactions \cite{LEP:2003aa}. These limits are often referred to as the LEP-II compositeness bounds. We impose constraints from measurements of the processes $e^+e^- \rightarrow e^+e^-, \, \mu^+ \mu^-,\, \tau^+ \tau^-$. The strongest constraints are therefore obtained when the mediator couples to electrons at tree-level, in which case couplings of order unity require a mediator heavier than $3 - 4 \,\,\mathrm{TeV}$. For mediators coupling primarily to $\mu$ and $\tau$ leptons at tree-level, we use the \runDM code~\cite{runDM} to evaluate the electron couplings at an energy scale of $209 \,\,\mathrm{GeV}$ and then apply the LEP-II constraints. While for axial-vector SM currents, this loop-induced coupling is small, it can be sizeable for vector currents, leading to limits of $m_V \gtrsim 300 \,\, \mathrm{GeV}$.

\paragraph{Collider Searches} LHC collisions probe leptophilic dark sectors through tree-level processes. Lepton pair production with a $V$ boson radiated by either of the particles in the final state give access to the dark sector. The reconstructed final state depends on the decay mode of the mediator. Drell-Yan processes pair produce charged leptons, and after $V$-radiation the final state can contain two leptons and missing energy or four leptons. The ATLAS four lepton cross section measurement~\cite{Aad:2014wra} constrain a possible mediator to be heavier than $m_V \sim 85 \,\,\mathrm{GeV}$, for $g_l = g_\chi = 1$ and for coupling to $e$ or $\mu$~\cite{Bell:2014tta}. For coupling to $\tau$ the measurements are not good enough to give any meaningful constraint. Conversely, for neutrino-charged lepton pair production through s-channel $W$ boson with the radiated mediator decaying invisibly, the final state contains one lepton and missing energy, as searched for in Refs.~\cite{ATLAS:2014wra,Khachatryan:2014tva}. The analysis in \Ref{delAguila:2014soa} found that for a leptophilic vector mediator the most stringent bounds come from processes with at least three leptons in the final states. All of these constraints are typically weaker than the LEP-II compositeness bounds~\cite{LEP:2003aa} and we therefore include only the latter in the phenomenological analysis performed in Sec.~\ref{sec:results}. However, we note that the relative importance of the different constraints depends on the assumed couplings and moving away from the case of $g_l = g_\chi = 1$ will shift the limits. 

Mono-jet searches, usually a powerful DM probe at colliders, are not well suited for models where the DM does not couple to quarks or gluons at tree-level. Nevertheless, there exist collider processes capable of probing these models. Initial state photon radiation at leptonic colliders may yield mono-photon events, that are bounded by LEP~\cite{Abdallah:2003np,Abdallah:2008aa}. For a mediator coupled to electrons with couplings of order one, the subsequent limits on the mediator mass are of order $m_V \gtrsim 400 - 500 \,\,\mathrm{GeV}$ for DM lighter than $\sim 100 \,\,\mathrm{GeV}$ \cite{Fox:2011fx}. Again, these limits are weaker than the LEP-II compositeness bound, though as above this statement holds only for $g_l = g_\chi = 1$.

\subsection{Bounds absent at tree-level}
\label{sec:absent}

\paragraph{EWPT} RG flow generates a mediator coupling to the Higgs current. Upon expanding this interaction around the vacuum
\be
\Delta \mathcal{L}_H = g_H \, H^\dagger i \overleftrightarrow D_\mu H \, V^\mu =
-  g_H \, \frac{2 c_w}{g_2} m_Z^2 \, Z_\mu V^\mu + \ldots  \ ,
\label{eq:massmixingoperator}
\ee
we identify a mass mixing between $V$ and the $Z$ boson. Here, $c_w$ and $g_2$ are the cosine of the weak mixing angle and the gauge coupling of the weak isospin gauge group, respectively. In the small mixing limit, this leads to a mixing angle between the two neutral vector particles
\be
\theta_{\rm mix} \simeq - g_H \, \frac{c_w}{g_2} \frac{m_Z^2}{m_V^2 - m_Z^2} \ . 
\ee
This mixing angle is bounded to be $\theta_{\rm mix} \lesssim 10^{-3}$ by ElectroWeak Precision Tests (EWPT)~\cite{Agashe:2014kda}. An approximate expression for $g_H$ is given in \Eq{eq:gHapprox}. We emphasize that the EWPT limits we impose are the most conservative ones, since we assume that the mass mixing is vanishing at $\Lambda_{\rm UV}$. This could easily not be the case in explicit microscopic realizations, and the mixing we account for is an irreducible contribution from loops of SM particles. Unless an unnatural cancellation takes place between UV and RG-induced terms, our analysis accounts for the minimum amount of mass mixing we expect in this class of models.

\paragraph{Dilepton Resonances at the LHC} The dilepton final state at the LHC is an excellent probe to look for new resonances. For the leptophilic models considered in this work, one may naively conclude that this channel is fruitless due to the lack of couplings to colored states. However, RG-induced couplings to quarks, although loop suppressed, still lead to meaningful bounds. The cross section in the narrow width limit reads
\be
\sigma_{p p \rightarrow l^+ l^-} = \frac{\pi \, BR_{V \rightarrow l^+ l^-}}{3 s}  
\sum_{q} C_{q\bar{q}} (m_V^2 / s) \, \left(g_{Vq}^2 + g_{Aq}^2\right) \ ,
\label{eq:sxLHCschannelNarrow}
\ee
where $BR_{V \rightarrow l^+ l^-}$ is the branching ratio of the decay $V \rightarrow l^+ l^-$. The parton luminosity $C_{q\bar{q}} (m_V^2 / s)$ for the quark $q$ reads
\be
C_{q\bar{q}}(y) = \int_{y}^1 d x \, \frac{f_{q}(x) \, f_{\bar{q}}(y / x) + f_{q}(y / x) \, f_{\bar{q}}(x)}{x}  \ ,
\ee
with $f_{q, \bar{q}}(x)$ the quark and antiquark parton distribution function (PDF). We use the PDFs from \Ref{Martin:2009iq} and we evaluate them using the public code available at the URL \href{https://mstwpdf.hepforge.org/}{https://mstwpdf.hepforge.org/}. The loop-induced couplings to quarks, whose approximate expressions can be found in \Eq{eq:onelooptoquarks}, are evaluated with \runDM. The cross section in \Eq{eq:sxLHCschannelNarrow} is then computed by evaluating the PDF and the couplings to quarks at the renormalization scale $m_V$. We impose the recent Run 2 bounds at $\sqrt{s} = 13 \, {\rm TeV}$~\cite{ATLAS:2017wce}\footnote{Note that the analysis of Ref.~\cite{ATLAS:2017wce} places limits on the cross section using data from both the di-electron and di-muon channels, with roughly equal constraining power from each. In our analysis, we assume a coupling only to muons (not electrons) and we therefore rescale the limit by a factor of $\sqrt{2}$ to account for the reduced statistical power which would arise from using only the di-muon channel. We also assume a constant signal acceptance of 40\%, which is typical for $Z'$ models in the di-muon channel \cite{ATLAS:2017wce}.}(for earlier studies by the ATLAS and CMS collaborations see Refs.~\cite{Aaboud:2016cth} and \cite{Khachatryan:2016zqb}, respectively). For tau pairs~\cite{Khachatryan:2016qkc}, RG-induced signals are not strong enough to put meaningful bounds.

\paragraph{Anomalous magnetic moments}

The coupling between the mediator and the lepton current in \Eq{eq:Jmuleptons} is responsible for a one-loop contribution to the anomalous magnetic moments~\cite{Agrawal:2014ufa}
\be\label{eq:g-2}
\delta(g-2)_l \simeq \frac{1}{12 \pi^2} \frac{m_l^2}{m_V^2} \left( g_{Vl}^2 - 5 g_{Al}^2 \right) \ .
\ee
The value of $(g-2)_\mu$ is measured to be larger than the Standard Model prediction at the $3\sigma$ level \cite{Beringer:1900zz}. With the contribution of vector couplings, the agreement between the theory and measurement can be improved and we therefore set limits by requiring that the total theoretical prediction does not exceed the measured value at the $2\sigma$ level. On the other hand, axial-vector couplings reduce the theory prediction, worsening the agreement. In this case, we require that the loop-induced contribution not exceed the $2\sigma$ uncertainty on the measured value. While determination of $(g-2)_e$ is more precise, the contribution from New Physics is expected to be smaller by a factor of $m_e^2/m_\mu^2 \approx 2 \times 10^{-5}$, leading to constraints on $m_V$ which are weaker by about a factor of 10 \cite{Bell:2014tta}. In contrast, experimental limits on $(g-2)_\tau$ are currently too weak to give meaningful constraints on New Physics \cite{Eidelman:2007sb}. While in principle there are also loop-induced contributions to $(g-2)_\mu$ from mediators coupling to $e$ or $\tau$, the corresponding limits are negligible in our case. This is why in Sec.~\ref{sec:results} we include only the $(g-2)_\mu$ bound, which applies to a mediator coupled to muons and generating an anomalous magnetic moment as given in \Eq{eq:g-2}.

\paragraph{Direct Detection} There is no direct detection at tree-level since the mediator couples only to leptons. However, RG mixing effects induce a coupling to light quarks~\cite{Crivellin:2014qxa,D'Eramo:2014aba}, which can be compactly written by using the effective Lagrangian in \Eq{eq:LagDD}. These loop-induced couplings can be evaluated by using the public code \runDM~\cite{runDM}. We provided analytical solutions for the low-energy couplings in Sec.~\ref{sec:RGE}, which are a very accurate approximation to the ones derived by performing the full RG evolution.  Previous works~\cite{Kopp:2009et,Bell:2014tta} focused on loop-induced direct detection rates for a mediator coupled to the lepton vector current. Here we point out a new effect, which follows from the general analysis of \Ref{D'Eramo:2014aba}. If the mediator couples to the axial-vector current, there is still a radiative contribution to direct detection rates arising from loops of the tau lepton (see \Eqs{eq:CVq}{eq:CAq}). Although suppressed by the tau's Yukawa coupling, it still leads to meaningful constraints. 

We present limits from the recent LUX WS2014-16 run \cite{Akerib:2016vxi} and projections for the upcoming LUX-ZEPLIN (LZ) experiment \cite{Akerib:2015cja}. Calculation of the approximate 95\% confidence limits from LUX is detailed in Appendix B of Ref.~\cite{Kavanagh:2016pyr} and calculation of the LZ projection is described in Appendix D of Ref.~\cite{D'Eramo:2016atc}, following the general procedure developed in Ref.~\cite{DelNobile:2013sia}.

\section{Results for single lepton flavor}
\label{sec:results}

\begin{figure*}[t!]
\centering
\hspace{0.75cm}\includegraphics[width=0.86\textwidth]{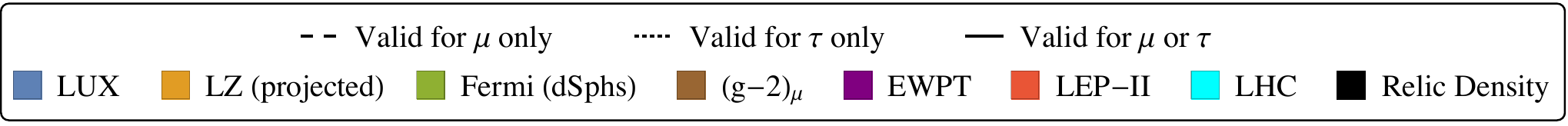}
{\includegraphics[trim={0.2cm 0.1cm 0.1cm 0.5cm},width=0.47\textwidth]{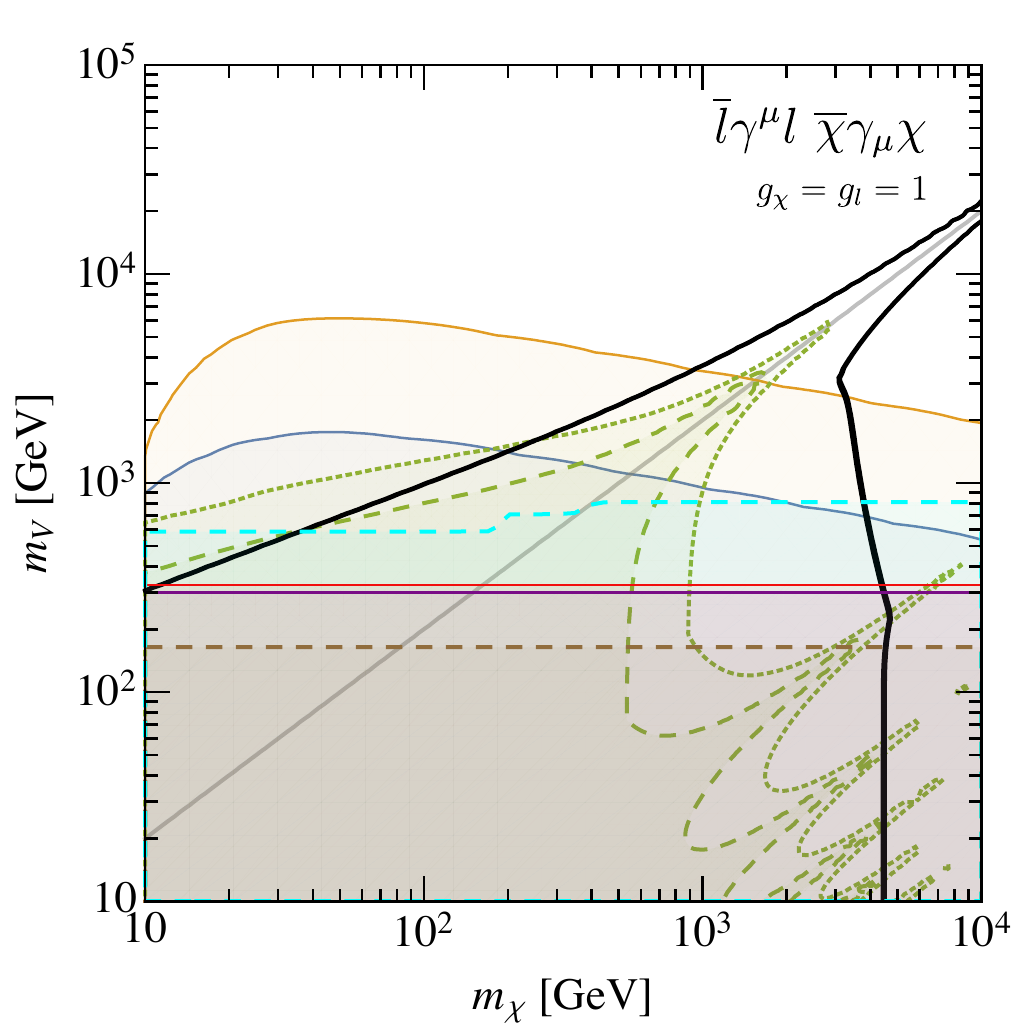}}
{\includegraphics[trim={0.2cm 0.1cm 0.1cm 0.5cm},width=0.47\textwidth]{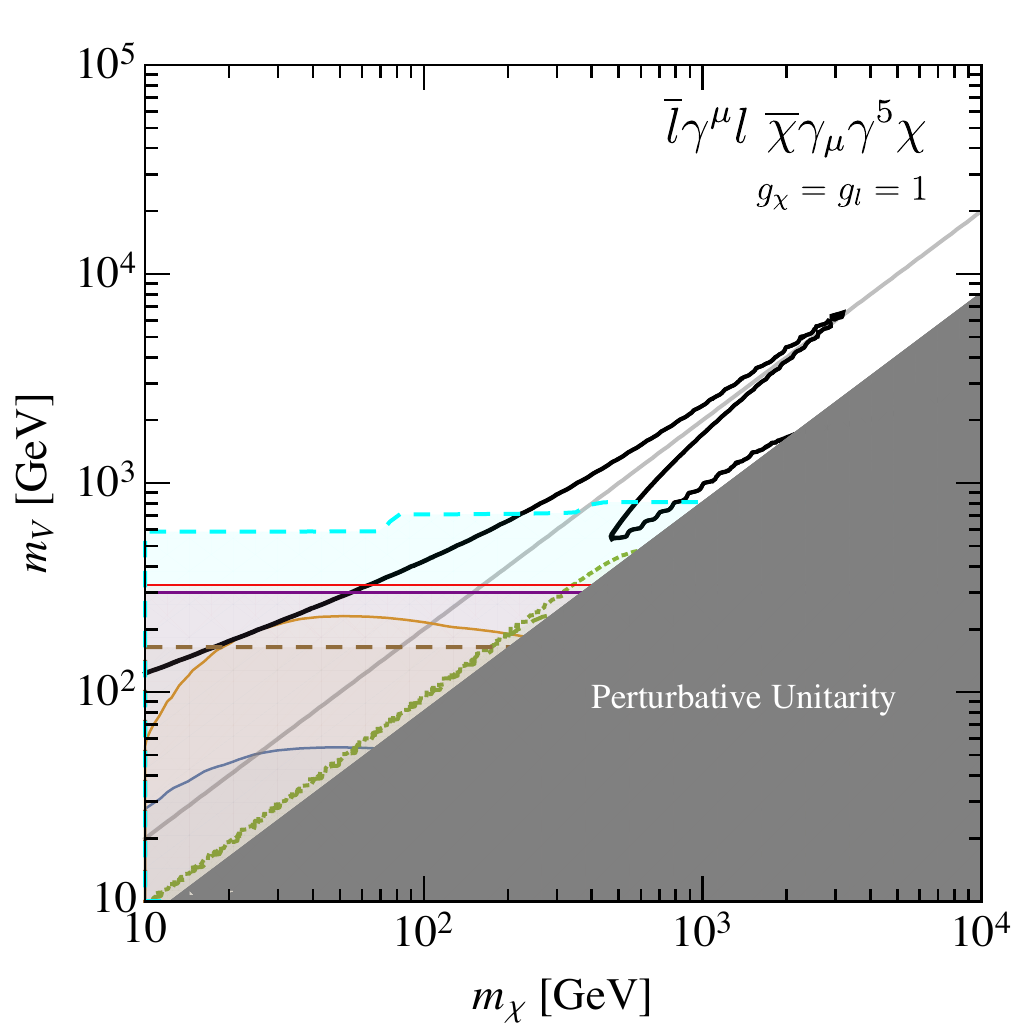}}
{\includegraphics[trim={0.2cm 0.1cm 0.1cm 0.5cm},width=0.47\textwidth]{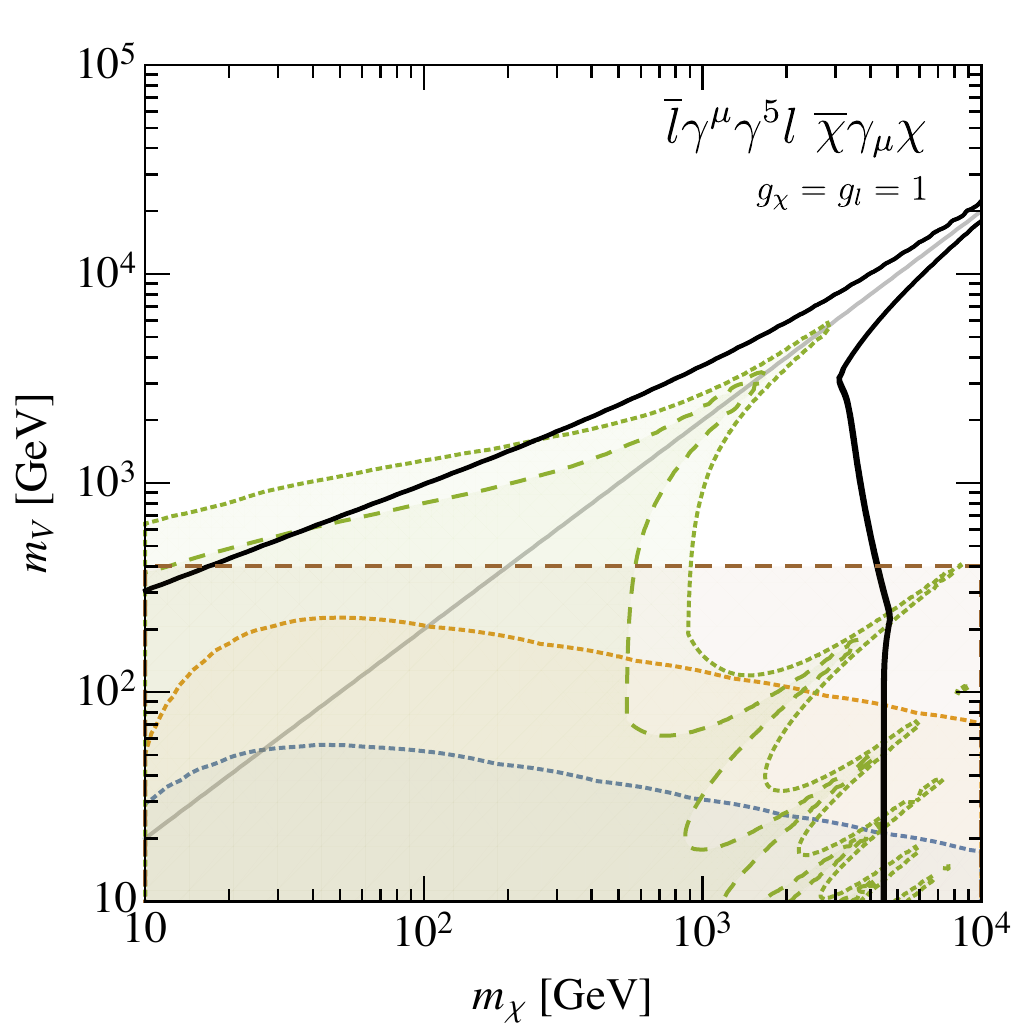}}
{\includegraphics[trim={0.2cm 0.1cm 0.1cm 0.5cm},width=0.47\textwidth]{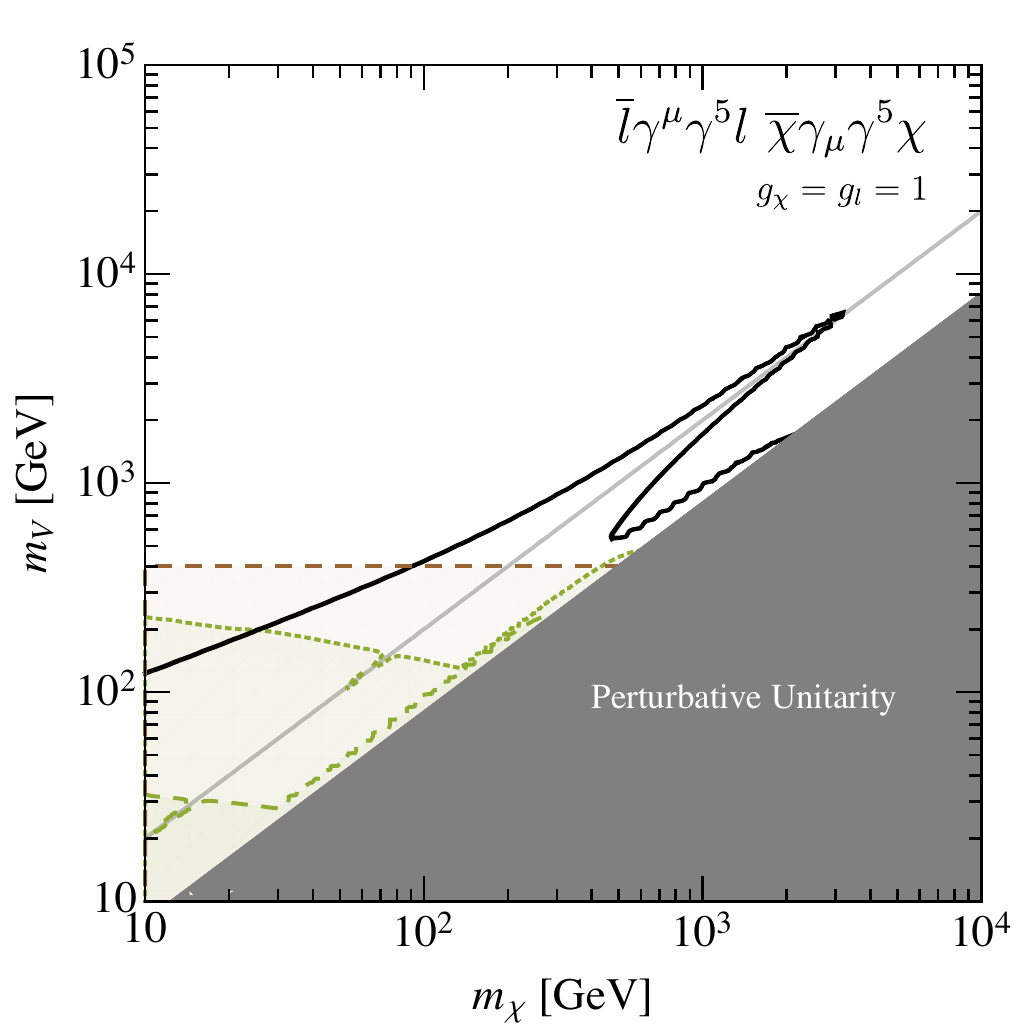}}
\caption{Constraints on leptophilic DM coupling to muons and taus through a heavy vector mediator, detailed in Sec.~\ref{sec:constraints}. Dashed (dotted) lines show constraints which are valid for mediators coupling only to $\mu$ ($\tau$) leptons. Solid lines are constraints valid in both cases. The solid grey area in the case of axial-vector couplings to DM is excluded as it violates perturbative unitarity~\cite{Kahlhoefer:2015bea}. The faint grey diagonal line is included to guide the eye and denotes the boundary $m_V = 2 m_\chi$. Note that a vector mediator coupling to electrons (limits not shown) is strongly constrained by the LEP-II  compositeness bound \cite{LEP:2003aa} for all types of interaction, with limits of $m_V \gtrsim \mathcal{O}(3-4 \,\,\mathrm{TeV})$. All constraints are at the 95\% confidence level. Note that for LZ we show projected (rather than current) constraints.}
\label{fig:Limits}
\end{figure*}

We analyze models where the vector mediator couples only to a single lepton flavor with dimensionless couplings of order one, namely $g_{\alpha\chi} = g_{\beta l} = 1$ (with $\alpha, \beta = V, A$). As emphasized at the end of Sec.~\ref{sec:simplifiedmodel}, the charge assignments are made at the simplified model cutoff $\Lambda_{\rm UV} = 10 \, {\rm TeV}$, and we assume no other interactions at such a scale. The case of a mediator coupled to electrons is not considered here, since the tree-level coupling to electrons is strongly constrained by the LEP-II compositeness bound. In this case, the mediator mass is constrained at the level of $m_V \gtrsim \mathcal{O}(3-4 \,\,\mathrm{TeV})$, which is much stronger than almost all other constraints and leaves only a small window of $m_\chi \gtrsim 1 \,\, \mathrm{TeV}$ in which DM could be produced by thermal freeze-out. 

Constraints for coupling to $\mu$ or $\tau$ are presented in the $(m_\chi, m_V)$ plane in Fig.~\ref{fig:Limits}. Here, dashed (dotted) lines show bounds that are valid only in the case of couplings to muons (taus). Solid lines are constraints which apply in either case. 

For a mediator having vector interactions with both DM and leptons (top-left panel), the dominant constraints come from direct and indirect detection. In the former case, this arises from a loop-induced interaction with quarks (and therefore nucleons) driven by the electromagnetic coupling, while the latter comes from tree-level annihilation to leptons. For couplings to muons, competitive constraints (particularly at high DM mass, $m_\chi \gtrsim 1 \,\,\mathrm{TeV}$) come from searches for dilepton resonances at the LHC. The same loop-induced couplings with quarks which lead to direct detection constraints allow for the hadronic production of the vector mediator, which in turn decays into leptons. Since we consider coupling to a single lepton flavor, decays to both charged lepton and the associated neutrino of the same flavor are always open. Mediator decays to DM pairs are only allowed for $m_V > 2 m_\chi$. Explicit expressions for the mediator width are provided in \ref{app:XS}. For couplings of order one as chosen in this Section, the mediator width is $6.6 \%$ ($4 \%$) if decays to DM pairs are (not) allowed. We impose the bounds from the recent analysis in \Ref{ATLAS:2017wce}, which provided limits on the production cross section for different mediator widths.  For coupling to muons, we obtain a limit of $m_V \gtrsim 700 \,\,\mathrm{GeV}$ for the mass of the mediator. Limits on heavy mediators decaying to taus are weaker, owing to the relative difficulty of reconstructing tau leptons in the detector, and no significant constraints can be drawn for coupling to taus. 

In the top-right panel we have the case of a mediator coupling to the vector current of leptons and the axial-vector current of DM. For couplings to muons, similar LHC constraints hold, which are the strongest ones. In contrast with the previous case, direct detection limits are much weaker, as the loop-induced interaction with nucleons is in this case velocity suppressed. Furthermore, indirect detection limits from Fermi are negligible as the annihilation cross section is $p$-wave suppressed (see \Eq{eq:sigma1leptons}). For a mediator coupled to taus, the strongest bounds (at the level of $m_V \gtrsim 300 \,\,\mathrm{GeV}$) come from EWPT and LEP-II compositeness bounds, which both arise from loop-induced couplings.
 
For axial-vector couplings to leptons and vector couplings to DM (bottom-left panel), the dominant constraint over much of the DM mass range is the Fermi limit on DM annihilation in dwarf spheroidal galaxies. In this case, the annihilation cross section is $s$-wave and proportional to the DM mass squared. At large DM masses, constraints on $(g-2)_\mu$ are also competitive when the mediator couples to muons. Direct detection proceeds through operator mixing. For axial-vector couplings to leptons, such a mixing is driven by the lepton Yukawa coupling, unlike the previous case of lepton vector current where it was driven by the EM coupling. As a result, these is no direct detection constraint for coupling to muons (and also no bounds from LHC and EWPT). In contrast, the tau Yukawa is large enough to induce an observable spin-independent direct detection signal (which is coherently enhanced), meaning that LUX can constrain such simplified models at the level of $m_V \gtrsim 50 \,\,\mathrm{GeV}$. This effect has not previously been pointed out for leptophilic DM.

Finally, for axial-vector couplings to both leptons and DM (bottom-right panel), the parameter space is more poorly constrained. Though $(g-2)_\mu$ constraints still apply for tree-level couplings to muons, the annihilation cross section is suppressed by the mass of the lepton in the final state, leading to weaker indirect detection limits. For couplings to the tau, the loop-induced direct detection cross section is a combination of velocity-suppressed spin-independent and unsuppressed spin-dependent interactions. The latter receives no coherent enhancement of the rate and therefore gives no appreciable bounds in this case.

\section{Hadronic contamination in indirect detection spectra}
\label{sec:contamination}

The gamma ray bounds in Sec.~\ref{sec:results} were obtained by considering tree-level processes. Radiative corrections can potentially alter the spectral features of the predicted flux of cosmic rays (CRs). As an example, Refs.~\cite{Kachelriess:2007aj,Bell:2008ey,Kachelriess:2009zy,Ciafaloni:2010qr,Ciafaloni:2010ti} considered leptophilic models and quantified these corrections for gamma rays arising from annihilations of DM particles with mass larger than the weak scale ($m_\chi \gtrsim m_{Z}$). In such a mass range, final state leptons have energies larger than the weak scale and the contamination of the spectra is due to electroweak bremsstrahlung. 

In this section we discuss contamination of CR spectra within our framework and for DM masses smaller than the weak scale  ($m_\chi \lesssim m_{Z}$). This effect is due to the mixing among the spin-one currents of SM particles coupled to the mediator~\cite{D'Eramo:2014aba}, and it was not pointed out in previous studies. More specifically to our leptophilic framework, RG flow generates a coupling of the mediator to the Higgs current through operator mixing (see the EWPT discussion in Sec.~\ref{sec:absent}). As a consequence, the DM acquires an effective vertex with the $Z$ boson, which opens up new channels for indirect detection spectra. For example, it may be possible to have DM annihilations with hadronic final states mediated by a virtual $Z$ boson exchanged in the s-channel. 

\begin{figure*}[t!]
\centering
\hspace{0.75cm}\includegraphics[width=0.9\textwidth]{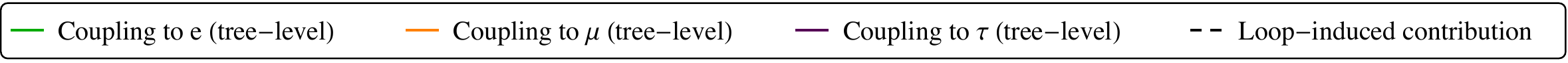}
{\includegraphics[trim={0.2cm 1.8cm 0.1cm 0.5cm},width=0.252\textwidth,clip]{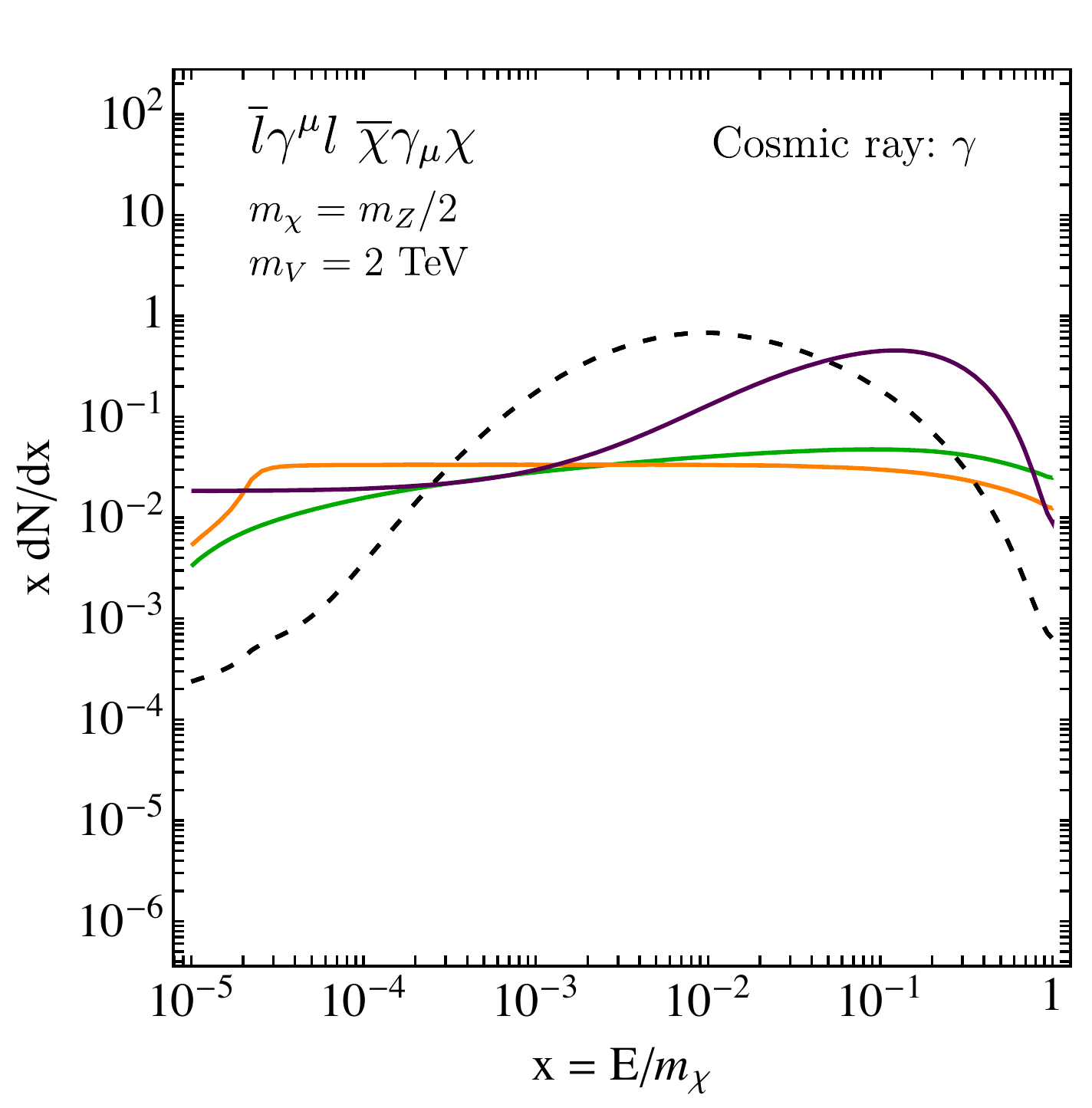}}
{\includegraphics[trim={0.9cm 1.8cm 0.1cm 0.5cm},width=0.24\textwidth,clip]{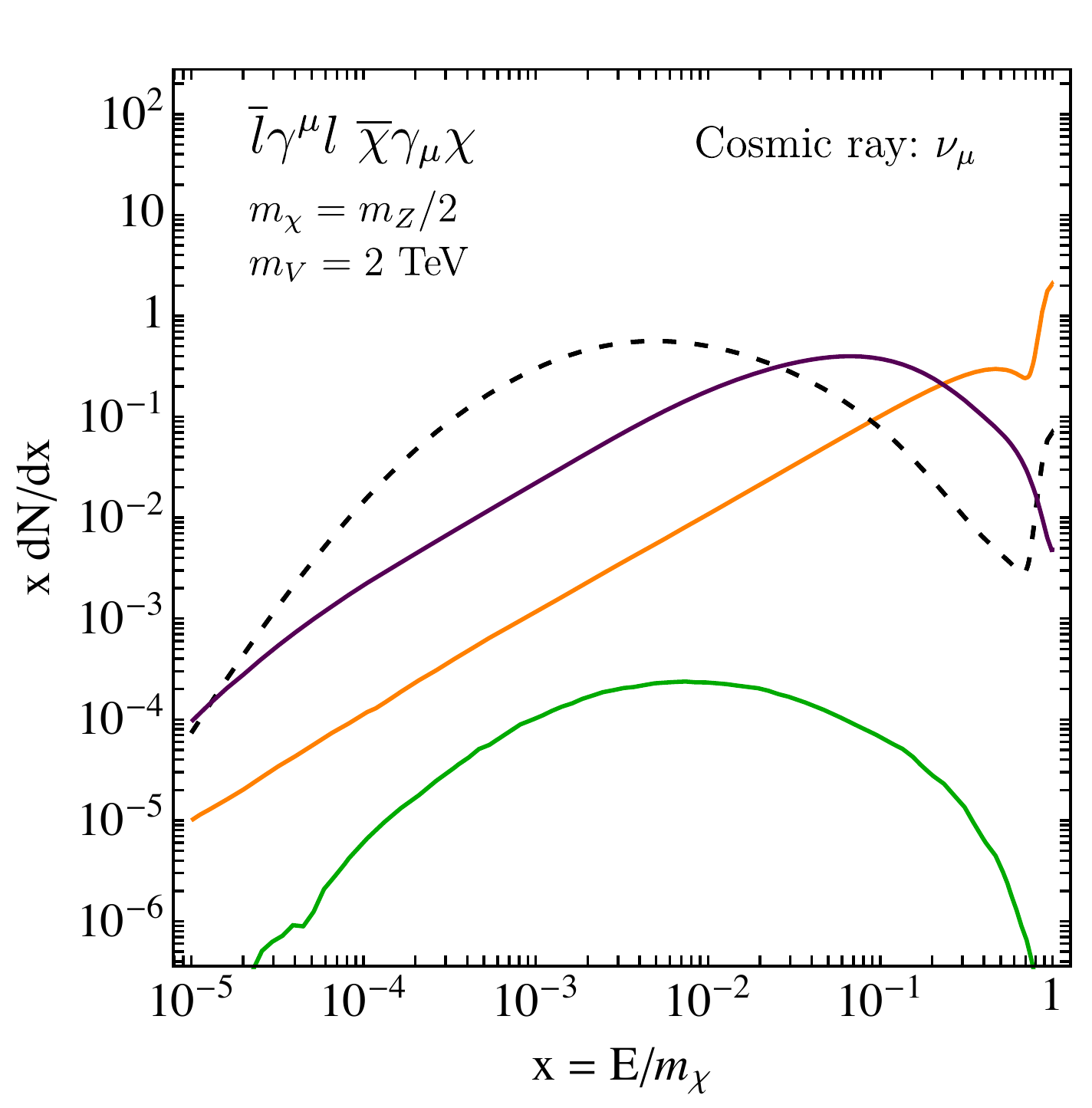}}
{\includegraphics[trim={0.9cm 1.8cm 0.1cm 0.5cm},width=0.24\textwidth,clip]{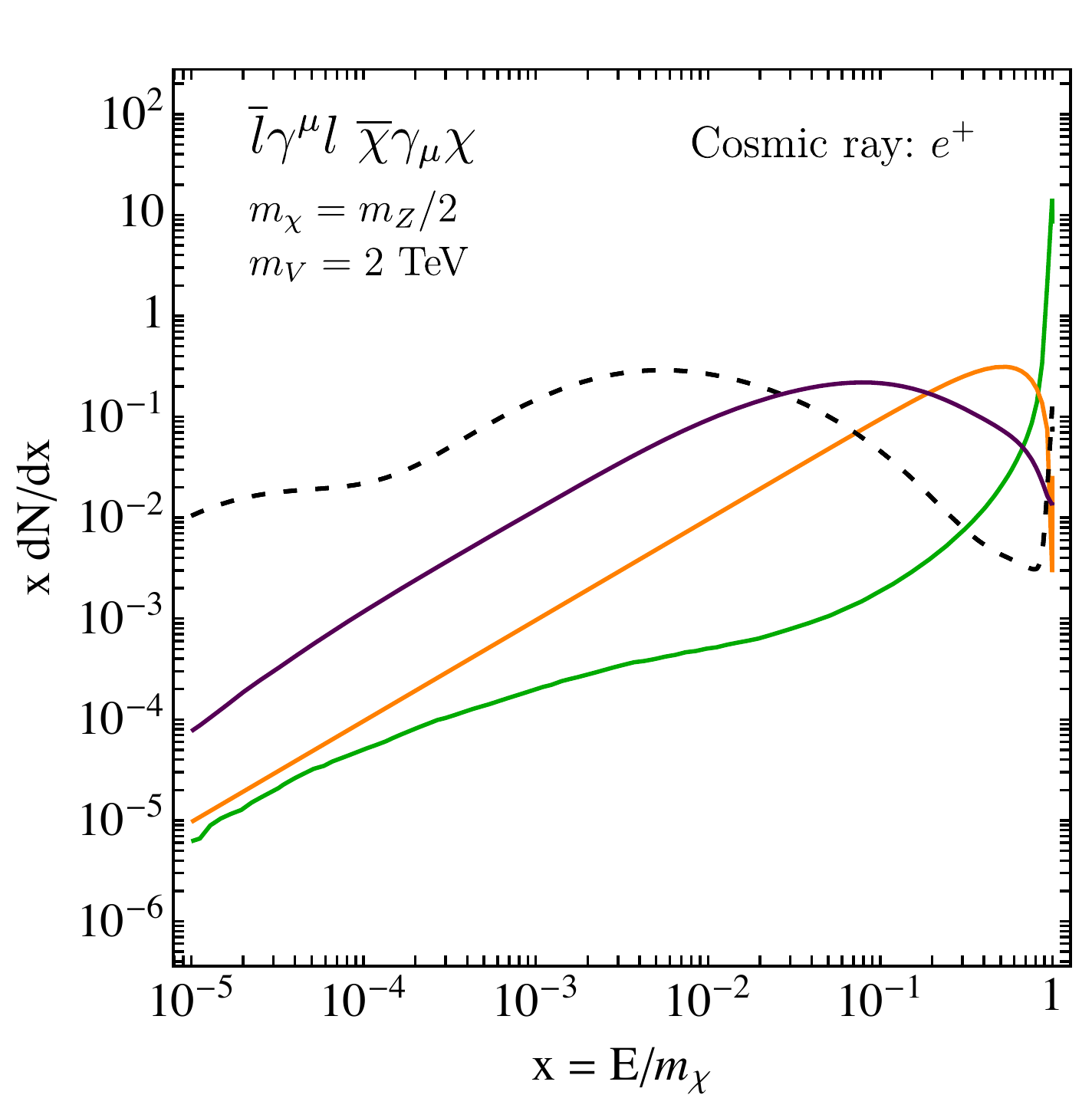}}
{\includegraphics[trim={0.9cm 1.8cm 0.1cm 0.5cm},width=0.24\textwidth,clip]{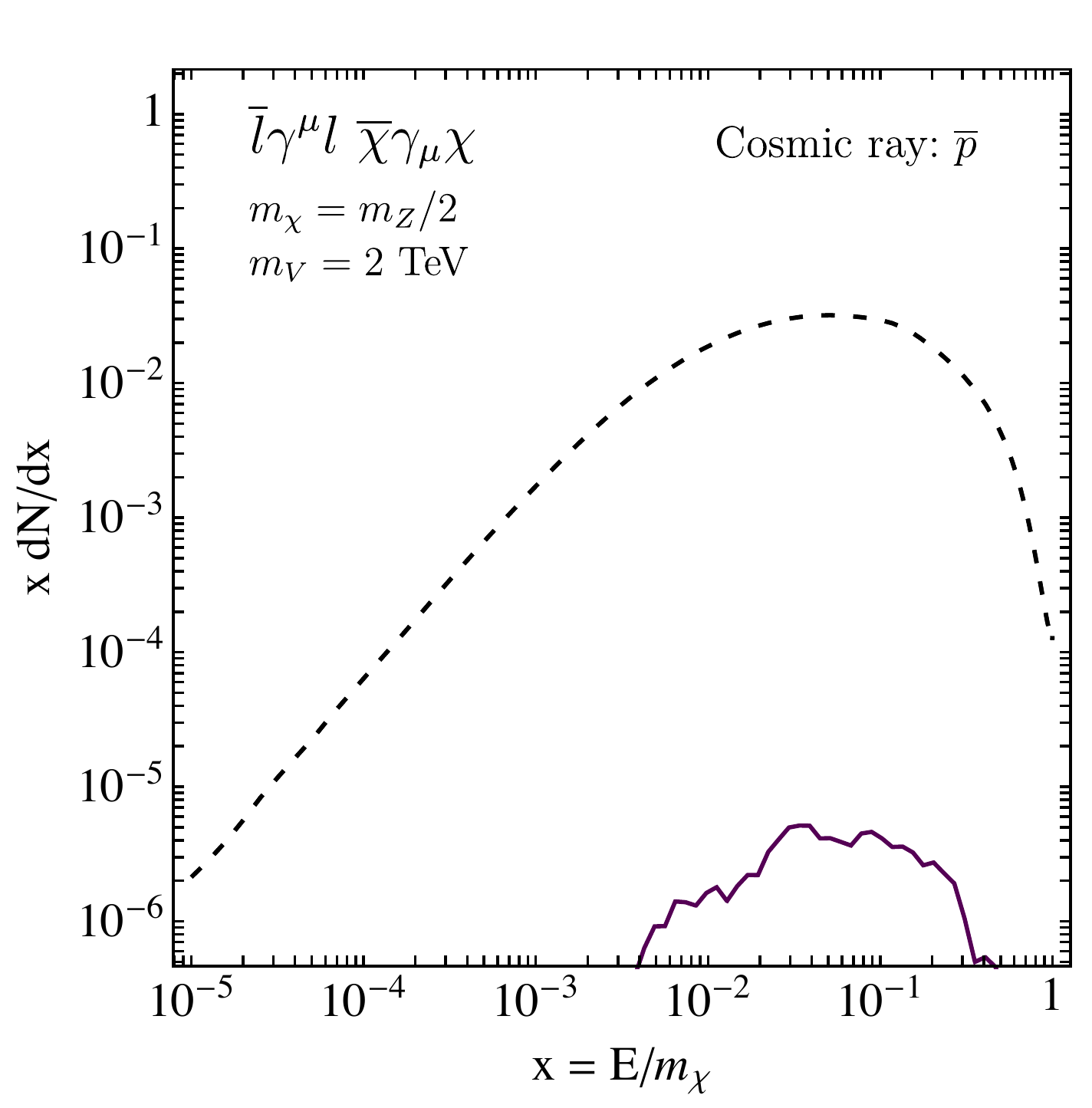}}
{\includegraphics[trim={0.2cm 0.1cm 0.1cm 0.8cm},width=0.252\textwidth,clip]{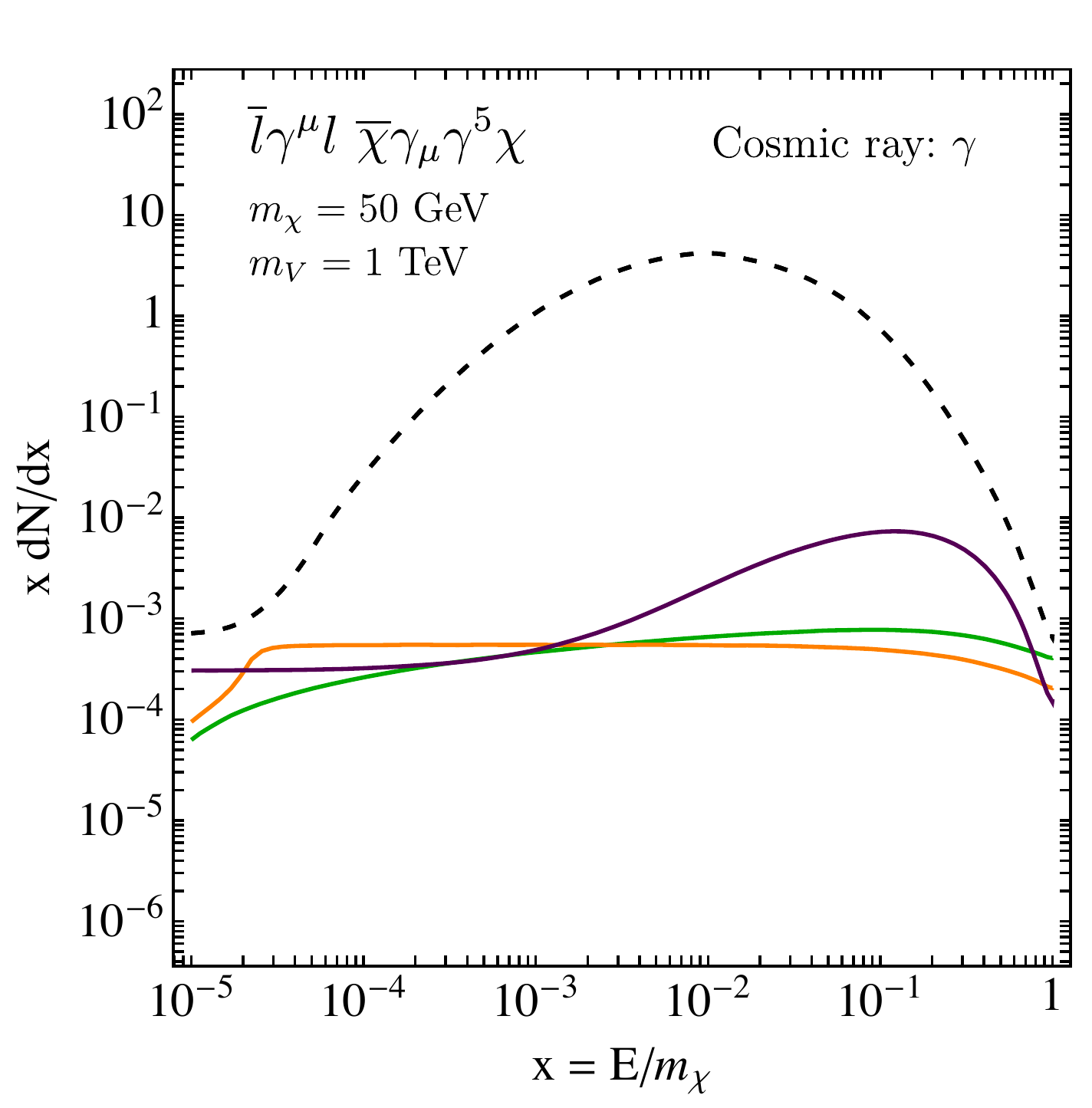}}
{\includegraphics[trim={0.9cm 0.1cm 0.1cm 0.8cm},width=0.24\textwidth,clip]{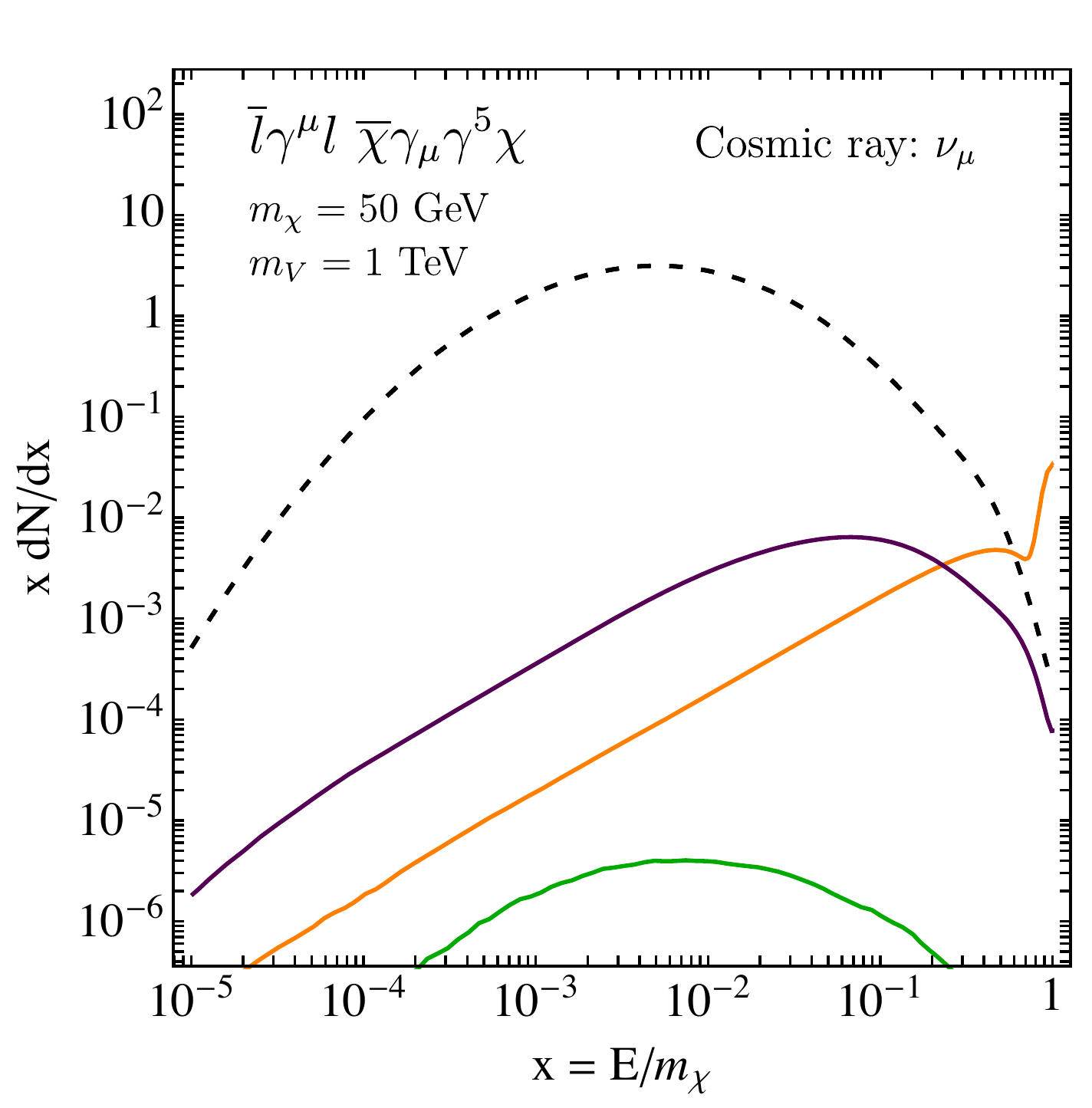}}
{\includegraphics[trim={0.9cm 0.1cm 0.1cm 0.8cm},width=0.24\textwidth,clip]{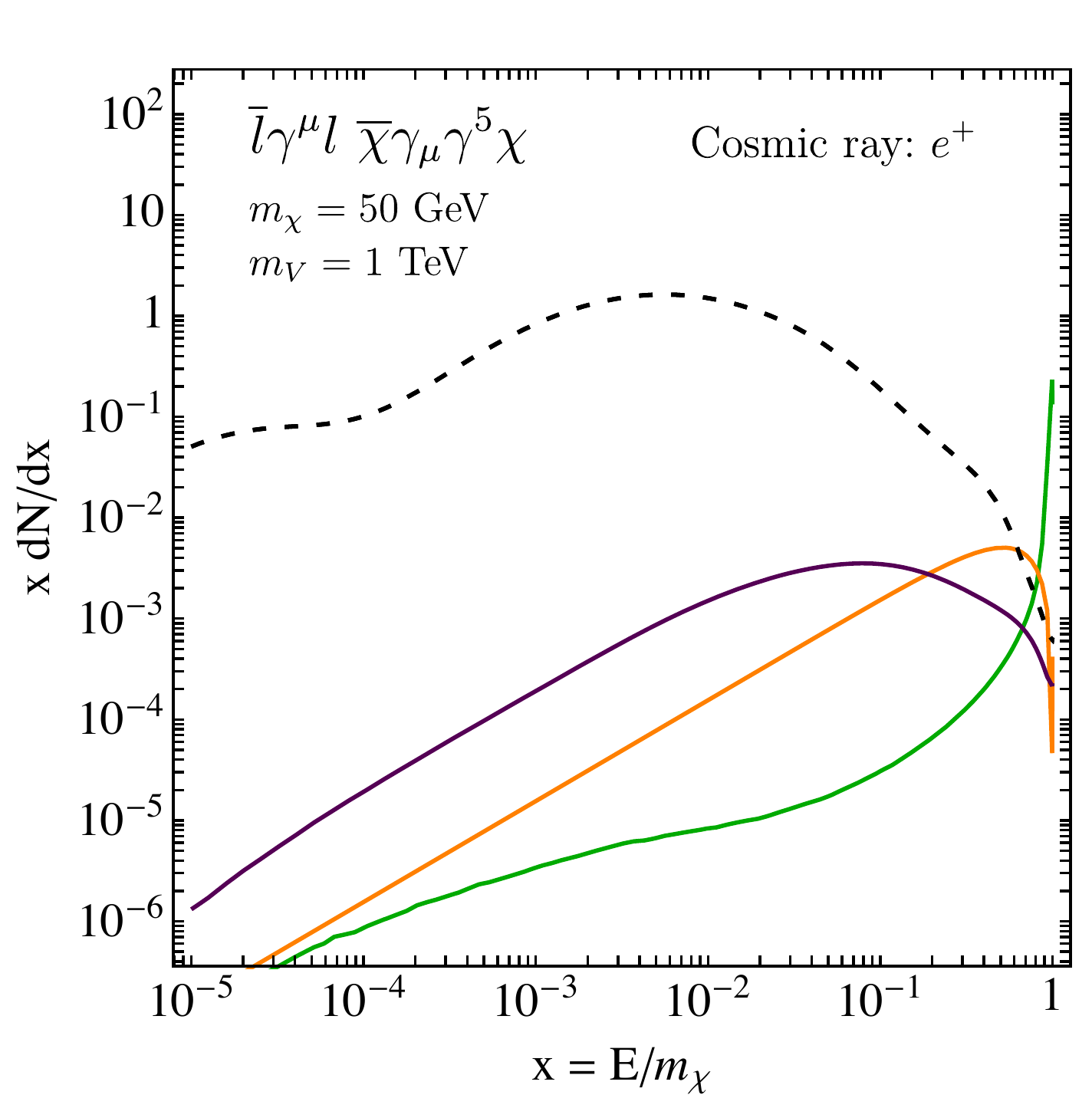}}
{\includegraphics[trim={0.9cm 0.1cm 0.1cm 0.8cm},width=0.24\textwidth,clip]{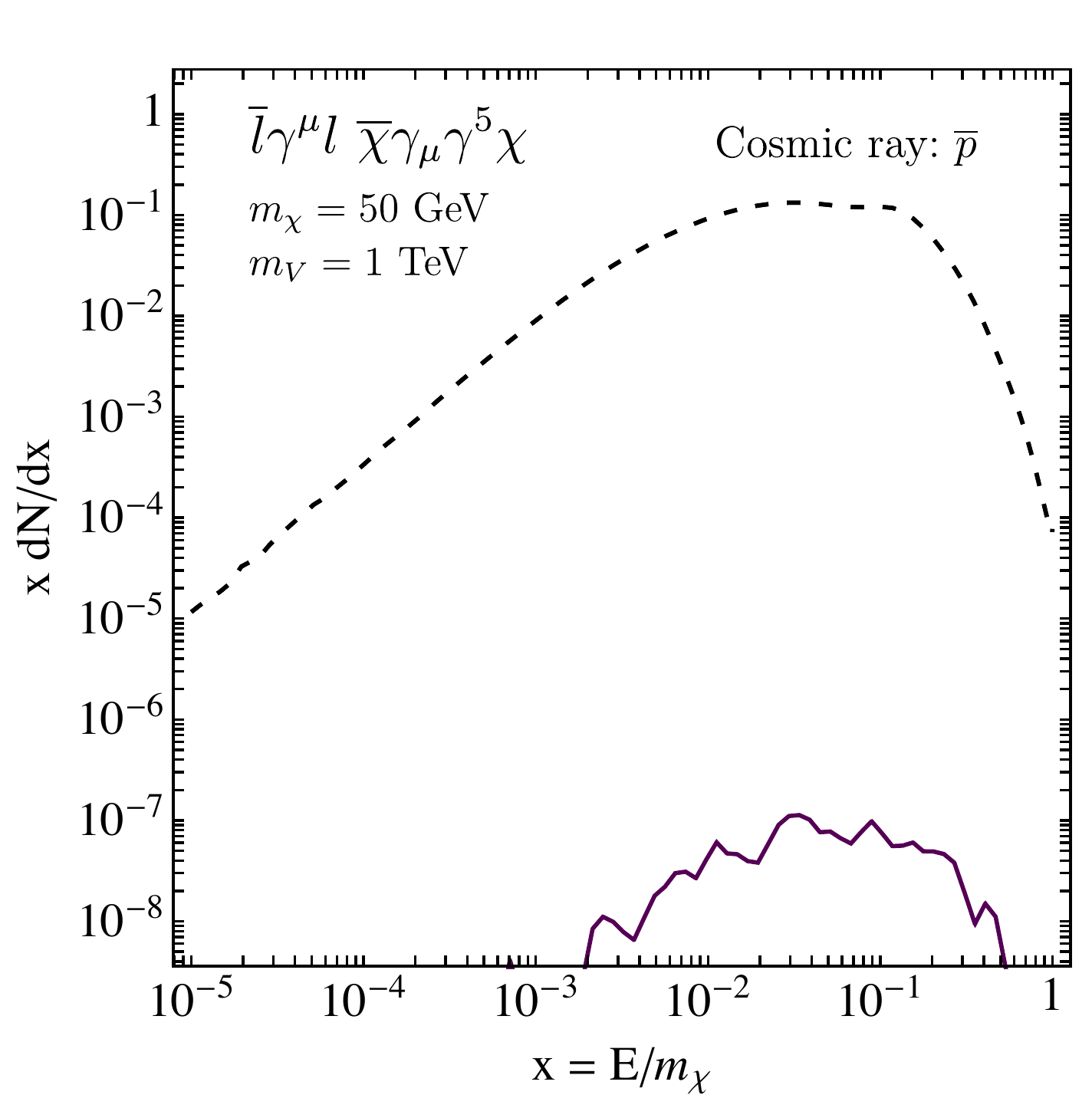}}
\caption{Spectrum of cosmic rays produced (per annihilation) for leptophilic DM. The top row shows results for vector interactions to both leptons and DM, while the bottom row shows results for vector couplings to leptons and axial-vector couplings to DM. Each column shows the spectrum of a different cosmic ray species (from left to right): $\gamma$, $\nu_\mu$, $e^+$ and $\overline{p}$. The tree-level contribution to the spectrum is shown as a solid colored line for a mediator coupling to electron, muon or tau leptons. The dashed line shows the loop-induced contribution. The total spectrum is the sum of the tree-level and loop-induced contributions.}
\label{fig:spectra}
\end{figure*}

The $Z$ boson can always be integrated out for direct detection scattering. However, for DM annihilation the center of mass energy is approximately twice the DM mass, meaning that we cannot always integrate out the $Z$. If the DM mass is close enough to the $Z$-resonance, we have to account for the full propagator of the intermediate $Z$ boson, and the resonant enhancement may give rise to hadronic final states overtaking the leptonic ones already present at the tree-level. The total CR spectrum per annihilation is in general the sum of the tree-level and loop-induced contributions. 
Explicitly, the primary spectra at production of  stable SM products $i$ can be written
\begin{align}\label{eq:TotaldNdx}
\begin{split}
\dbd{N_i}{x} =  \left(  \frac{\sigma_{\chi \chi \rightarrow Z' \rightarrow l^+ l^-}}{\sigma_\mathrm{tot}} \dbd{N_i^l}{x} + \sum_f  \frac{\sigma_{\chi \chi \rightarrow Z \rightarrow f\overline{f}}}{\sigma_\mathrm{tot}} \dbd{N_i^f}{x} \right) \ ,
\end{split}
\end{align}
where $x=E/m_\chi$ and $\sigma_{\mathrm{tot}}=\sigma_{\chi \chi \rightarrow Z' \rightarrow l^+ l^-} + \sum_f  \sigma_{\chi \chi \rightarrow Z \rightarrow f\overline{f}} $ is the total annihilation cross section. Here ${\rm d}N_i^{l(f)}/{\rm d} x$ is the spectrum of products $i$ arising solely from the primary $l(f)$ with energy $E$, obtained from \textsc{PPPC4DMID} \cite{Cirelli:2010xx,Ciafaloni:2010ti}.

In Fig.~\ref{fig:spectra}, we show the spectra at production of CRs from DM annihilation. We consider four different cases of stable SM products (different columns), and we take two benchmark scenarios: vector couplings to both leptons and DM (top row) and vector couplings to leptons with axial-vector coupling to DM (bottom row). For the former scenario, we show the special case where the DM can annihilate resonantly through a $Z$ boson exchanged in the s-channel, $| m_\chi - m_Z / 2 | \ll \Gamma_Z$. We plot the tree-level CR spectra (first term in the parentheses of Eq.~\eqref{eq:TotaldNdx}) assuming annihilation to electrons, muons or taus (solid colored lines). We also plot the contribution to the spectrum arising from $Z$-mediated annihilation (dashed line), computed by summing the spectra for all possible final state annihilation products (second term in the parentheses of Eq.~\eqref{eq:TotaldNdx}).

For vector couplings to both DM and leptons (top row), mixing with the $Z$-boson leads to annihilation into quarks, which produces, via parton showering and hadronization, the broad low-energy bump in the CR spectra. In fact, in this case both the tree-level and loop-induced annihilations are $s$-wave processes, meaning that annihilation to quarks is loop-suppressed and so generally subdominant to annihilation to leptons. However, for DM masses around $m_\chi \sim m_Z/2$, $Z$-mediated annihilation to quarks is resonantly enhanced and can therefore become dominant at low energies (as in Fig.~\ref{fig:spectra}). The gamma-ray flux (top left-most panel) receives a substantial contribution from parton showering and hadronization of quarks. This means that the indirect detection limits presented in Fig.~\ref{fig:Limits} should in principle be modified close to the resonance ($m_\chi \approx m_Z/2$)  to account for this enhanced flux. Such an enhancement may be relevant for  Simplified Model fits to the Galactic Centre Excess, where a DM particle with masses in the range 30 - 70 GeV has been proposed (see e.g.~Refs.~\cite{Alves:2014yha,Abdullah:2014lla,Calore:2014nla}). We note however that such a Simplified Model is likely to be strongly constrained by direct detection (as described in Sec.~\ref{sec:results}).

\begin{figure}[t!]
\centering
{\includegraphics[width=0.4\textwidth]{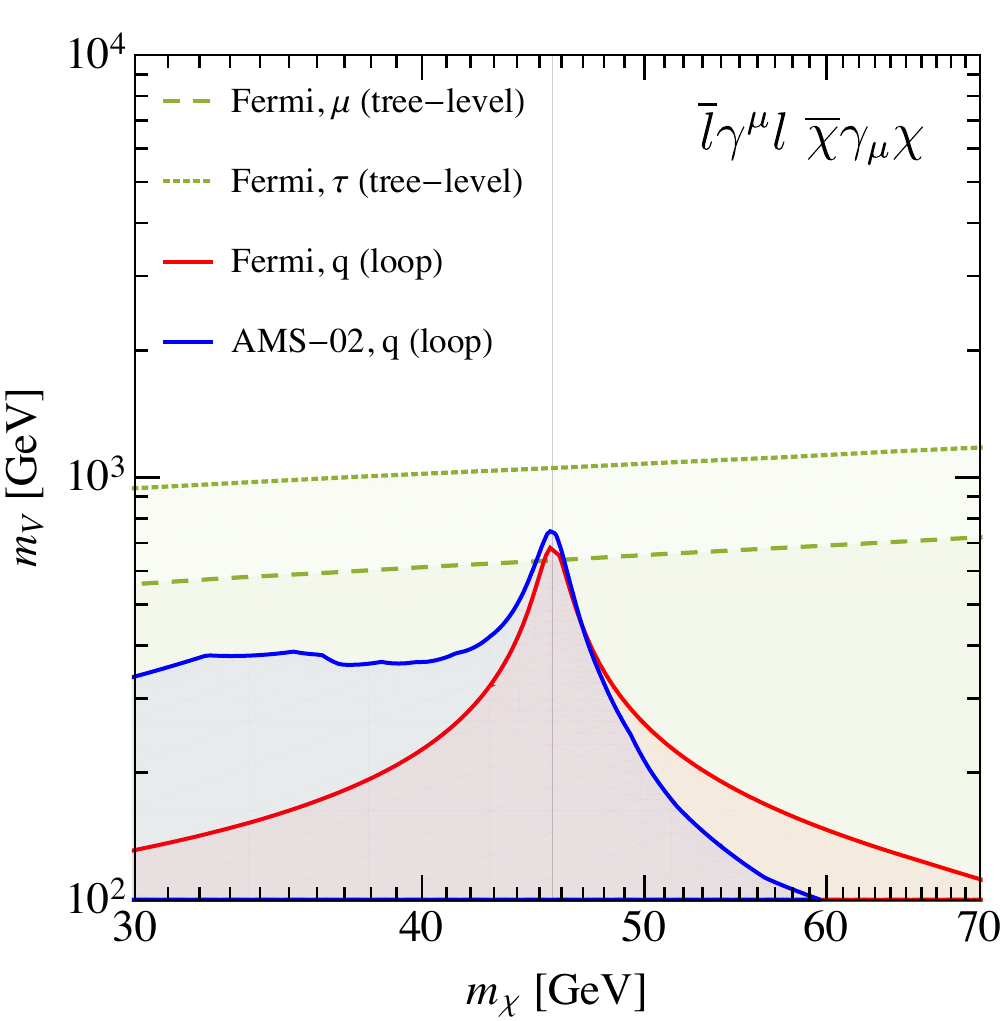}}
\caption{Indirect detection constraints on leptophilic DM coupling to mu and tau leptons through the interaction $\overline{l}\gamma_\mu l \, \overline{\chi} \gamma^\mu \chi$. We show the tree-level constraints from Fermi dSphs arising from this tree-level coupling to muons and taus (dashed and dotted green, respectively). We also show the limits from loop-induced $Z$-mediated annihilation to quarks: gamma ray bounds from Fermi dSphs (solid red) and antiproton bounds from AMS-02 (solid blue).}
\label{fig:contamination}
\end{figure}

Loop-induced enhancements to the indirect detection spectra are particularly noticeable for anti-protons (top right-most panel). At the tree-level, annihilation to leptons produces very few anti-protons.\footnote{The anti-proton flux coming from the annihilation either to electrons or muons is zero for $m_\chi < m_{Z}$. For $m_\chi \gg m_{Z}$ one can get a sizable amount of anti-protons via electroweak gauge bosons bremsstrahlung~\cite{Kachelriess:2007aj,Bell:2008ey,Kachelriess:2009zy,Ciafaloni:2010qr,Ciafaloni:2010ti}.} On the contrary, the anti-proton flux from (loop-induced and resonant) annihilation to quarks is some 3 orders of magnitude larger. It is natural to investigate whether the anti-proton flux measured by AMS-02~\cite{Aguilar:2016kjl,Cuoco:2016eej} can put bounds on leptophilic models. Assuming for simplicity universal anti-proton spectra for all DM annihilations to quarks (and taking them to be the same as for b quarks), we answer this question in Fig.~\ref{fig:contamination}. For a DM mass at the $Z$ resonance, the loop-induced anti-proton flux leads to constraints on the mediator mass at the level of $m_V \gtrsim 750\,\,\mathrm{GeV}$. Remarkably, this is competitive with constraints from the Fermi gamma-ray search in dwarf Spheroidals already plotted in Fig.~\ref{fig:Limits}, which arise from tree-level, leptonic annihilations. 

For vector couplings to leptons and axial-vector couplings to DM (bottom row), tree-level annihilation is $p$-wave suppressed. Mixing with the $Z$ leads to an effective axial-vector coupling to leptons, which gives rise to $s$-wave annihilation (see Eq.~\ref{eq:sigma0leptons}). Thus, the loop contribution can be substantially larger than the tree-level one, though the main contribution in this case is from annihilation to $b$ quarks only (as the cross section is now suppressed by the mass of the final state particle). In this scenario, the loop-induced correction is generally expected to be large compared to the tree-level spectrum, regardless of the DM mass. However, we note that even including the large loop contribution, the CR fluxes are not large enough to be constrained by current indirect detection searches (limits from Fermi are negligible in the top right panel of Fig.~\ref{fig:Limits}). 

We do not consider the impact of hadronic contamination for axial-vector couplings to leptons. As already mentioned in Sec.~\ref{sec:results}, the mixing for axial-vector currents is typically smaller than for the vector current, being driven now by the tau Yukawa. Contamination of the spectra through $Z$-mediated annihilation is therefore typically small in that scenario.

\section{Conclusions}
\label{sec:conclusions}

WIMPs remain among the best motivated candidates for particle DM. Models where the DM particle is coupled only to SM leptons alleviate the tension with experimental constraints, since the most severe bounds come from experimental processes involving hadrons. However, hadronic processes are still able to probe {\it leptophilic} dark sectors due to the RG-evolution of the couplings with the energy scale. More precisely, radiative effects generate couplings to quarks at higher-orders, which although suppressed still give sizeable signals. 

This work exploited the relevance of these effects for leptophilic dark sectors with a vector portal. We considered the massive mediator coupled to only a single lepton flavor at a time, and in each case we imposed the experimental bounds listed in Sec.~\ref{sec:constraints}. They arise both from tree-level and radiatively-induced processes. For a mediator coupled to electrons, the LEP-II compositeness bound is extremely severe. For comparable couplings to DM and electrons, it easily overtakes bounds from DM searches. For this reason, we have focused only on the cases of coupling to $\mu$ and $\tau$. 

Our main results are shown in Fig.~\ref{fig:Limits}. In the top panels we show the cases of vector couplings to muons and taus. A noteworthy feature of our results is that the most important constraints come from hadronic processes: direct detection in the case of vector couplings to DM and LHC dilepton resonances in the case of axial-vector couplings to DM (for which the direct detection rate is suppressed). In the bottom panels we show the cases of axial-vector couplings to leptons. In contrast, the dominant constraints here are from leptonic processes: indirect detection in the case of taus and $(g-2)_\mu$ in the case of muons. Only for axial-vector coupling to taus and vector couplings to DM do we find an appreciable hadronic constraint. In this case, mixing effects driven by the tau Yukawa lead to spin-independent DM-nucleon scattering, which is strongly constrained by LUX and other direct detection experiments. This effect was not pointed out in previous studies of leptophilic DM. 

Indirect detection is the only canonical WIMP search with tree-level signals in leptophilic models. We studied higher order corrections to cosmic ray spectra in Sec.~\ref{sec:contamination}. While it was known that for high DM mass ($m_\chi \gtrsim m_Z$) electroweak bremsstrahlung may affect indirect detection spectra, no previous studies pointed out such a contamination in the low mass region. For vector coupling to leptons, RG flow induces an effective coupling of the DM particle to the $Z$ boson, which in turn mediates annihilation to hadrons. For a DM mass sufficiently close to the $Z$ resonance, these radiatively-induced contributions to the CRs fluxes may dramatically alter the predicted spectra. This is illustrated in Fig.~\ref{fig:spectra}, where for both vector and axial-vector DM couplings we compare the tree-level and loop-induced spectra of four different stable SM products. Final state anti-protons are a particular interesting case. At low DM masses ($m_\chi \lesssim m_Z$), the tree-level anti-proton flux is either zero (coupling to electron and muon) or extremely suppressed (coupling to tau), thus the radiatively-induced contribution to the spectrum is significant. We illustrate this in Fig.~\ref{fig:contamination}, where we superimpose the recent AMS anti-proton bounds over the other constraints used in this work. The bounds are competitive for a DM mass close enough to the $Z$ resonance ($m_\chi \simeq m_Z/2$).  

Leptophilic dark sectors are an attractive scenario. Motivated by anomalies in the observed CRs fluxes, they also alleviate the tension with experimental bounds from DM searches involving hadronic processes. In spite of expectations based on tree-level calculations, the constraints are not as mild when accounting for radiative corrections. This work quantified the magnitude of these effects and put novel bounds on the allowed parameter space. We also highlighted new prospects for probing leptophilic dark sector, such as future EWPT and dilepton resonance searches at hadron colliders. We find it intriguing that a leptophilic DM particle may ultimately be discovered in experimental searches not only probing dark sector couplings to leptons but also to quarks. 

\section*{Acknowledgements}

We thank Stefano Profumo for his valuable feedback on the manuscript and the Journal Referee for her/his very insightful and constructive comments. FD acknowledges very helpful discussions with Mike Hance  about dilepton resonance searches. FD is supported by the U.S. Department of Energy grant number DE-SC0010107 (FD). BJK is supported by the European Research Council ({\sc Erc}) under the EU Seventh Framework Programme (FP7/2007-2013)/{\sc Erc} Starting Grant (agreement n.\ 278234 --- `{\sc NewDark}' project).

\appendix

\section{Decay Widths and Cross Sections}
\label{app:XS}

We collect results for the mediator partial decay widths and non-relativistic DM annihilation cross sections. The mediator width for the decay to a fermion/antifermion pair is
\be
\Gamma_{V \; \rightarrow \bar{f} f} = \frac{g_{Vf}^2 + g_{Af}^2}{12 \pi} m_V \ .
\ee
Here, $g_{Vf}$ and $g_{Af}$ are the mediator couplings to the vector and axial-vector fermion currents, respectively. 

We expand DM annihilation cross sections in partial waves, $\sigma_{\chi \chi \rightarrow {\rm final}} v_{\rm rel} = \sigma^{(s)}_{\chi \chi \rightarrow {\rm final}} +
\sigma^{(p)}_{\chi \chi \rightarrow {\rm final}} v_{\rm rel}^2$.  Annihilations to leptons, both charged and neutrinos, are always kinematically allowed 
\begin{align}
\label{eq:sigma0leptons} \sigma^{(s)}_{\chi \chi \rightarrow l^+ l^-} = & \,
\frac{g_{V \chi}^2  \, (g_{Vf}^2 + g_{Af}^2)}{\pi} 
\frac{m_\chi^2}{(4 m_\chi^2 - m_V^2)^2} + \\ & \nonumber
\frac{g_{A \chi}^2  \, g_{Af}^2}{2 \pi} \frac{m_l^2}{m_V^4} \left( 1 - \frac{m_l^2}{m_\chi^2}\right)^{1/2}  \ , \\
\label{eq:sigma1leptons} \sigma^{(p)}_{\chi \chi \rightarrow l^+ l^-}  = & \, \frac{g_{A \chi}^2  \, 
(g_{Vf}^2 + g_{Af}^2)}{6 \pi} \frac{m_\chi^2}{(4 m_\chi^2 - m_V^2)^2} \ .
\end{align}
Models with a vector (axial-vector) coupling to the DM have an s-(p-)wave cross section surviving the $m_l \rightarrow 0$ limit. If we keep terms suppressed by the lepton mass $m_l$, we also have a s-wave piece for theories where the couplings are both axial-vector.  DM annihilations to mediators, if allowed, can be important since they are s-wave processes
\be
\sigma^{(s)}_{\chi \chi \rightarrow V V} = \frac{(g_{V\chi}^4 - 6 g_{V\chi}^2 g_{A\chi}^2 + g_{A\chi}^4) + \frac{8 g_{V\chi}^2 g_{A\chi}^2}{\epsilon_V^2} }{16 \pi \, m_\chi^2} \, f_{VV}(\epsilon_V)   \ ,
\ee
where $\epsilon_V \equiv m_V / m_\chi$ and $f_{VV}(x) = (1 -x^2)^{3/2} (1 - x^2 / 2)^{-2}$.



\bibliography{VectorLeptophilic}

\end{document}